\documentclass[%
 aip,
 amsmath,amssymb,
 reprint,%
]{revtex4-1}

\usepackage{graphicx}
\usepackage{dcolumn}
\usepackage{bm}

\usepackage[utf8]{inputenc}
\usepackage[T1]{fontenc}
\usepackage{mathptmx}
\usepackage{etoolbox}

\usepackage{subcaption}
\usepackage{hyperref}

\makeatletter
\def\@email#1#2{%
 \endgroup
 \patchcmd{\titleblock@produce}
  {\frontmatter@RRAPformat}
  {\frontmatter@RRAPformat{\produce@RRAP{*#1\href{mailto:#2}{#2}}}\frontmatter@RRAPformat}
  {}{}
}%
\makeatother
\begin{document}

\preprint{AIP/123-QED}

\title[FreeMHD: validation and verification of the open-source, multi-domain, multi-phase solver for electrically conductive flows]{FreeMHD: validation and verification of the open-source, multi-domain, multi-phase solver for electrically conductive flows}

\author{Brian Wynne}
\affiliation{Department of Mechanical and Aerospace Engineering, Princeton University, Princeton, NJ 08540, US}

\author{Francisco Saenz}
\affiliation{Department of Mechanical and Aerospace Engineering, Princeton University, Princeton, NJ 08540, US}

\author{Jabir Al-Salami}
\affiliation{ 
Research Institute for Applied Mechanics, Kyushu University, Fukuoka, Japan}%

\author{Yufan Xu}
\affiliation{Princeton Plasma Physics Laboratory, Princeton, NJ 08540, US}

\author{Zhen Sun}
\affiliation{Princeton Plasma Physics Laboratory, Princeton, NJ 08540, US}

\author{Changhong Hu}
\affiliation{ 
Research Institute for Applied Mechanics, Kyushu University, Fukuoka, Japan}%

\author{Kazuaki Hanada}
\affiliation{ 
Research Institute for Applied Mechanics, Kyushu University, Fukuoka, Japan}%

\author{Egemen Kolemen}
    \email{ekolemen@pppl.gov}
 \homepage{https://control.princeton.edu/}
\affiliation{Department of Mechanical and Aerospace Engineering, Princeton University, Princeton, NJ 08540, US}
\affiliation{Princeton Plasma Physics Laboratory, Princeton, NJ 08540, US}%


\date{28 July 2024}

\begin{abstract}
The extreme heat fluxes in the divertor region of tokamaks may require an alternative to solid plasma-facing components, for the extraction of heat and the protection of the surrounding walls. Flowing liquid metals are proposed as an alternative, but raise additional challenges that require investigation and numerical simulations. Free surface designs are desirable for plasma-facing components (PFCs), but steady flow profiles and surface stability must be ensured to limit undesirable interactions with the plasma. Previous studies have mainly used steady-state, 2D, or simplified models for internal flows and have not been able to adequately model free-surface liquid metal (LM) experiments. Therefore, FreeMHD has been recently developed as an open-source magnetohydrodynamics (MHD) solver for free-surface electrically conductive flows subject to a strong external magnetic field. The FreeMHD solver computes incompressible free-surface flows with multi-region coupling for the investigation of MHD phenomena involving fluid and solid domains. The model utilizes the finite-volume OpenFOAM framework under the low magnetic Reynolds number approximation. FreeMHD is validated using analytical solutions for the velocity profiles of closed channel flows with various Hartmann numbers and wall conductance ratios. Next, experimental measurements are then used to verify FreeMHD, through a series of cases involving dam breaking, 3D magnetic fields, and free-surface LM flows. These results demonstrate that FreeMHD is a reliable tool for the design of LM systems under free surface conditions at the reactor scale. Furthermore, it is flexible, computationally inexpensive, and can be used to solve fully 3D transient MHD flows.

\end{abstract}

\maketitle

\section{Introduction}



A challenge remaining in the pursuit of fusion energy is handling large heat and particle fluxes. These can be especially damaging to the interior of the reactor, specifically in the divertor region. \cite{pitts2019physics}
An alternative is to use a liquid metal (LM) to provide a self-healing and replenishing surface. A LM with a free surface could serve as the plasma-facing component, and offer advantages such as providing heat transfer through convection, as well as eliminating erosion and the need for component replacement. 
Additionally, some LM designs can offer low hydrogen recycling regimes, meaning that the excess fuel encountering the surface stays in the LM instead of returning to the plasma \cite{kaita2019fusion}. Lithium, for example, has been shown to reduce tritium recycling, with a key benefit being an increase in plasma performance \cite{krasheninnikov2003lithium,boyle2017observation}.

However, along with the advantages come additional challenges that must be solved. 
One such challenge is reducing the opposing Lorentz force, which acts as a magnetohydrodynamic (MHD) drag to cause a pileup of flow across a magnetic field \cite{sun2023magnetohydrodynamics}.  
Other challenges involve flow separation, droplet ejection, and keeping the surface as flat as possible to avoid surface deformities which would collect a large heat flux from the exhaust plasma at a shallow angle in the divertor, leading to concerns of evaporation or splashing \cite{nygren2004fusion}. 

In order to address these challenges, a key step is the development of numerical solvers that can simulate the behavior of conductive flows across magnetic fields. The primary motivation is to develop reliable computational code that can accurately predict and replicate real flows under laboratory conditions, thereby enabling simulations under reactor regimes and informing the design of future fusion experiments. 
Existing challenges include the computational complexity and memory-intensive nature of simulating multi-phase, three-dimensional flows, and electric currents with free surfaces. 



Previous studies have mainly used steady-state, 2D, or simplified models for internal flows and have not been able to adequately model free-surface LM experiments.
A 2D model was developed for studying a magnetic field gradient for film flow\cite{gao2002numerical}. This model predicted the negative effects of flows across these fields such as hydraulic jump-like patterns forming, but did not include all the effects involving recirculating currents. 
Subsequently, a 3D code was developed for free surface flows using induced-magnetic field equations and confirmed the need for 3D codes to capture phenomena that could not be described by 2D simulations\cite{huang20023d}. 
Simulations using HyPerComp Incompressible MHD solver for Arbitrary Geometries (HIMAG) were developed for predicting and modeling the effects of the magnetic environment on the flow characteristics of the LM flows using the electric potential formulation \cite{morley2004progress}. 
However, HIMAG is not open-source or as flexible as OpenFOAM based solvers.
Recently, a set of incompressible MHD flow solvers have been developed, with one solver for single–phase flow in multiple electro–coupled domains and one solver for two-phase flows \cite{siriano2024multi}. However, these solvers are currently separate and not yet integrated. Additionally, the verification focuses on closed channel analytical results and validation using bubbles, so the solvers have not been shown to be validated against free surface bulk-flowing liquid metal experiments. Therefore, an open-source computational model capable of accurately simulating 3D free-surface liquid metal flows in the presence of applied magnetic fields is needed, as previous examples lack flexibility, open-source availability, and the ability to adequately model relevant plasma-facing component experiments.

In response to this need, MHD OpenFOAM solvers have been developed \cite{AlSalami2022numerical} and have now been formalized and called {FreeMHD}\footnote{FreeMHD \url{https://github.com/PlasmaControl/FreeMHD}.} as an open-source, custom simulations to computes multi-region, free-surface flows in magnetic fields. 
FreeMHD is a necessary tool for the fusion community, offering the ability to solve design and implementation problems for LM PFC research. Moreover, it is open-source, readily available for use, and fully parallelized for CPUs. 
FreeMHD is a comprehensive solver for free surface liquid MHD flows and can be used for any flow with MHD conditions.


The objective of this study is to describe FreeMHD, present the validation and verification, and explain the utility of the computational solver for magnetohydrodynamics and fusion-related applications. 
This paper seeks to describe the motivation and need for these new developments, provide details on the model, and explain how it is implemented. It is the contention of this paper that the open-source nature of FreeMHD enables future development, allowing further improvements and additional features to be added. Throughout, it will be shown that the validation of FreeMHD with experiments has been successful and that FreeMHD provides a fast, accurate, and accessible way to model free-surface liquid metal flows under fusion-relevant conditions. This enables the planning and testing of divertor concepts in tokamak reactors.

Section \ref{section:MaterialMethods} discusses the mathematical model, while Section  \ref{section:NumericallMethods} describes the numerical method and solver specifications. 
Section \ref{section:Results} presents the results of validation and verification, and Section \ref{section:Discussion} provides a discussion of the results and applications. Lastly, Section \ref{section:Conclusions} summarizes and suggests future directions and development.

\section{Mathematical Models}
\label{section:MaterialMethods}

\subsection{Governing Equations}

The modeling of electrically conductive flow begins with using the incompressible, Eulerian form of the Navier-Stokes Equations 
\begin{equation}
    \frac{\partial \rho}{\partial t}+\nabla \cdot (\rho \mathbf{U})=0
    \label{eqn:cont}
\end{equation}
\begin{equation}
    \frac{\partial \rho \mathbf{U}}{\partial t} +\nabla \cdot (\rho \mathbf{U} \otimes \mathbf{U}) = -\nabla p + \nabla \cdot \boldsymbol{\underline{\underline{\tau}}} +\mathbf{F_g}+\mathbf{F_{ST}}+\mathbf{F_L}
    \label{eqn:navstokes}
\end{equation}
for the conservation of mass and momentum, with parameters defined in Table \ref{tab:params}.

An alternative pressure was defined for numerical convenience, $p'=p-\rho (\mathbf{g} \cdot \mathbf{x})$. 
The viscous stress tensor is represented by $\boldsymbol{\underline{\underline{\tau}}}=\mu\left(\nabla \mathbf{U} +(\nabla \mathbf{U})^\mathrm{T} \right)$

\begin{table}[t]
    \small
    \centering
    \caption{Parameter Definitions}
    {
        \begin{tabular}{c|cc}
        \hline
        \textbf{Symbol} & \textbf{Unit} & \textbf{Description}  \\
        \hline
        $\mathbf{U}$            & $\mathrm{m/s}$& Fluid Velocity  \\
 $\mathbf{B}$& T&Magnetic Field\\
 $\mathbf{J}$& $\mathrm{A/m^2}$&Electric Current Density\\
        $\rho$                  & $\mathrm{kg/m^3}$& Fluid Density   \\
        $p$                     & $\mathrm{Pa}$& Fluid Pressure  \\
        $p'$                     & $\mathrm{Pa}$& Alternative Fluid Pressure  \\
        $\boldsymbol{\underline{\underline{\tau}}}$     & $\mathrm{Pa}$& Viscous Stress Tensor   \\
        $\mathbf{F_g}$          & $\mathrm{N/m^3}$& Gravity Volume Force     \\
        $\mathbf{F_b}$          & $\mathrm{N/m^3}$& Surface Tension Volume Force     \\
        $\mathbf{F_b}$          & $\mathrm{N/m^3}$& Lorentz Volume Force     \\
        $\mathbf{g}$            & $\mathrm{m^2/s}$& Gravity Vector  \\
        $\mathbf{x}$            & $\mathrm{m}$& Position Vector \\
 $\mathbf{n}$& $\mathrm{-}$&Normal Vector\\
 $\mathbf{\alpha}$& $\mathrm{-}$&Volume Fraction\\
 $D$& $\mathrm{m^3}$&Volume Domain\\
 $\mathbf{\kappa}$& $\mathrm{1/m}$&Curvature Vector\\

        $\mathbf{\gamma_{st}}$ &$\mathrm{N/m}$& Fluid Surface Tension Coefficient \\
        $\sigma$                     & $\mathrm{S/m}$& Fluid Electrical Conductivity  \\
        $\sigma_w$                     & $\mathrm{S/m}$& Solid Wall Electrical Conductivity  \\
 $\mu_0$& $\mathrm{N/A^2}$&Vacuum permeability\\
 $U$& $\mathrm{m/s}$&Fluid Velocity Magnitude\\
        $B$                     & $\mathrm{T}$& Magnetic Field Strength  \\
        $L$                     & $\mathrm{m}$& Channel Length  \\
        $\ell$                     & $\mathrm{m}$& Channel Half Width  \\
        $h$                     & $\mathrm{m}$& Fluid Height  \\
        $\mathbf{S}_f$& $\mathrm{-}$&  Surface Normal Vector\\
        $N_f$                     & $\mathrm{-}$& Number of Faces\\
        $\Delta t$                     & $\mathrm{s}$& Time Step\\
 $\Delta x$& m&Cell Length\\
        $V_C$                     & $\mathrm{m^3}$& Cell Volume  \\

        \hline
        \end{tabular}
    }
    \label{tab:params}
\end{table}

The finite volume OpenFOAM framework with the volume of fluid (VoF) method is used for interface capturing, where a function $I(\mathbf{x},t)$ is defined at each point and is equal to 1 or 0 to indicate fluid 1 or fluid 2, over the volume domain $D$. Then,  the volume fraction in a control volume $D_i$ within $D$ is  defined as 
\begin{equation}
    \alpha_i=\frac{1}{|D_i|}\int_{D_i} I(\mathbf{x},t) dV
\end{equation}
over the domain. The evolution of the volume fraction for an incompressible flow is described by 
\begin{equation}
    \frac{\partial \alpha}{\partial t}+\nabla\cdot (\alpha \mathbf{U}) =0
\end{equation}
The physical properties of the fluid in each cell are calculated from the volume fraction using 
\begin{equation}
    \xi=\alpha\xi_1 +(1-\alpha)\xi_2
\end{equation}

where $\xi_1$ and $\xi_2$ are the fluid properties of density, viscosity, and electrical conductivity. Lastly, the surface tension force between the gas-liquid interface is modeled using the continuum surface force (CSF) method \cite{brackbill1992continuum}.  Here, the force due to surface tension is calculated as $\mathbf{F_{st}}=\gamma_{st} \boldsymbol{\kappa} \nabla \alpha$, and has been included as a body force in the momentum balance. Here, $\boldsymbol{\kappa}=\nabla \cdot \left( \frac{\nabla \alpha}{|\nabla \alpha|}\right)$ is the curvature vector between the fluid-fluid interface and $\gamma_{st}$ is the fluid surface tension coefficient.

The magnetic Reynolds number is defined as $Re_m = \mu_0 \sigma U\ell$, for characteristic values of velocity $U$, length $\ell$, and fluid conductivity $\sigma$, where $\mu_0$ is the vacuum permeability. 

$Re_m $ represents the ratio of convection to diffusion of the magnetic field, and for $ Re_m  \ll 1$, diffusion is dominated and the induced magnetic field due to induced currents can generally be neglected when compared to the applied magnetic field $\mathrm{B}$ \cite{moreau1990magnetohydrodynamics}.  
The result is that the magnetic field is essentially unaffected by the flow, and instead of a magnetic field being coupled with the velocity field, one can just consider the velocity field coupled to the background magnetic field.

This is known as the inductionless approximation and allows the electric field to be expressed as the gradient of the scalar electric potential ($\mathbf{E}=-\nabla \phi$  )
\cite{muller2001magnetofluiddynamics}. Then from Ohm's law, 
\begin{equation}
    \mathbf{J} = \sigma (-\nabla \phi+ (\mathbf{U}\times \mathbf{B}))
    \label{eqn:Ohms}
\end{equation}
and conservation of current density, 
 \begin{equation}
    \nabla \cdot \mathbf{J} = 0
    \label{eqn:CurrCons}
\end{equation}
the electric potential Poisson equation is expressed as 
\begin{equation}
    \nabla^2 (\sigma \phi) = \nabla \cdot \left((\sigma  \mathbf{U}) \times \mathbf{B} \right)
    \label{eqn:potEPoisson}
\end{equation}

This is discretized for each face $f$ 
\begin{equation}
    \sum_{f=1}^{N_f} \left[  \sigma \frac{\partial \phi}{\partial \mathbf{n}}S \right]_f = \sum_{f=1}^{N_f} \left[  \sigma (\mathbf{U}\times \mathbf{B}) \cdot  \mathbf{S} \right]_f
    \label{eqn:potEPoissonDiscretized}
\end{equation}
for $N_f$ number of faces, and is then used to solve for the electric potential $\phi$.

The electric current density at the cell faces is calculated using variables evaluated at each face center, 
\begin{equation}
    J_{n_f} = \left[ 
    \sigma \left(- \frac{\partial \phi}{\partial \mathbf{n}} S + (\mathbf{U}\times \mathbf{B}) \cdot \mathbf{n}
    \right)
    \right]_f
    \label{eqn:Jfaces}
\end{equation}
Then the Lorentz force ($F_L$) is calculated using its conservative form and reconstructed to the cell center for each cell $C$ with 
\begin{equation}
    (J\times B)_C = - \frac{1}{V_C} \sum_{f=1}^{N_f} J_{n_f} (\mathbf{B} \times \mathbf{x}S)_f - \mathbf{x}_C \times \left[ \frac{1}{V_C} \sum_{f=1}^{N_f} J_{n_f} (\mathbf{B} S)_f
    \right]
    \label{eqn:JxBConservative}
\end{equation}
following a conservative scheme for calculating the Lorentz force   \cite{ni2007current}.


\subsection{Boundary conditions}

The boundary conditions involve flow variables of velocity and pressure as well as electric potential and must be defined for all boundaries such as inlets, outlets, openings to the atmosphere, and interfaces between fluid and solid regions. 
The flow variable boundary condition at the inlet can be generally set through a fixed velocity, fixed pressure, or a fully developed velocity profile.
For a set velocity, the pressure at the inlet is then set to a zero gradient. For the outlet, a Dirichlet boundary condition is typically used for pressure at a reference value. 
For the top opening, the total pressure is fixed, allowing gas flow in or out. 
At the interface of fluid and solid walls, the gas and liquid boundary conditions involve the no-slip velocity condition  ($\mathbf{U}=0$). Additionally, considering the volume fraction for the free surface, a fixed contact angle is set for surface tension effects at the wall. 
The pressure at the wall is defined by a Neumann boundary condition, with the pressure gradient $ \partial p' / \partial \mathbf{n} $, such that the flux on the boundary is specified by the velocity condition, here the no-slip walls, where $\mathbf{n}$ is the wall normal vector. 
Electric current conservation at the inlet and outlet can be imposed by ensuring no electric current enters or exits through these boundaries, by setting $\partial \phi / \partial \mathbf{n} = (\mathbf{U} \times \mathbf{B}) \cdot \mathbf{n}$. 
When applying an external current density, the electric potential gradient at the boundary was set based on $\nabla \phi = \pm I_0 / (A \sigma_w)$, where $I_0$ is the applied current in Amperes, $A$ is the area of the boundary in meters, and $\sigma_w$ is the electric conductivity of the solid wall in siemens per meter.

\section{Numerical Methods}
\label{section:NumericallMethods}

\begin{table}[h!]
\centering
\small 
\begin{tabular}{|>{\centering\arraybackslash}m{1.4cm}|>{\centering\arraybackslash}m{1.4cm}|>{\centering\arraybackslash}m{1.8cm}|>{\centering\arraybackslash}m{1.8cm}|>{\centering\arraybackslash}m{1.5cm}|}
\hline
\textbf{Material} & \textbf{Density, \( \rho \) (kg/m\(^3\))} & \textbf{Kinematic Viscosity, \( \nu \) (m\(^2\)/s)} & \textbf{Electrical Conductivity, \( \sigma \) (S/m)}  &\textbf{Surface Tension, \( \sigma_{st} \) (N/m)}\\
\hline
Galinstan & 6360 & $2.98 \times 10^{-7}$ & $3.10 \times 10^6$  &0.533\\
\hline
NaK & 868.2 & $1.05 \times 10^{-6}$ & $2.879 \times 10^6$  & --- \\ 
\hline
\end{tabular}
\caption{Properties of Liquid Metals: Galinstan (eutectic gallium-indium-tin alloy, 67\% Ga, 20.5\% In, 12.5\% Sn at 20$^{\circ}$C)  \cite{morley2003mtor} and NaK (eutectic sodium-potassium alloy, 22\%Na 78\%K at 20$^{\circ}$C)  \cite{ODonnell1989thermophysical}  }
\label{eqn:tab}
\end{table}

\subsection{Solver Specifications}
The primary electric potential solver of FreeMHD is called \texttt{epotMultiRegionInterfoam}, which is discretized with implicit Euler time-stepping. Despite its first-order accuracy, this scheme’s stability allows for taking large time steps, which helps carry out efficient simulations of transient phenomena. 

FreeMHD has been tested and used in parallel and is typically run with CPUs on Princeton's Stellar and Della clusters. Table \ref{tab:cpuhours} contains examples of the CPU Hours and time to run on the cluster for the Shercliff, Dam Breaking, and LMX-U Cases. Additionally, the typical cell size and maximum time step are listed for reference. 

\begin{table}[]
\begin{tabular}{l|llll}
\hline
\textbf{Case}                                                                           & \textbf{CPU Hours} & \textbf{Time to run} & \textbf{Cell size} & \textbf{Time Step} \\ \hline
\textit{\textbf{Closed Channel}}                                                        & 12                 & 1 hr                 & 1-1000$\mu$m           & 0.01-1ms               \\ \hline

\textit{\textbf{\begin{tabular}[c]{@{}l@{}}Free Surface, \\ Dam Breaking\end{tabular}}} & 47.3               & 1.3 hr               & 500$\mu$m              & 1 ms                   \\ \hline

\textit{\textbf{\begin{tabular}[c]{@{}l@{}}Free Surface, \\ LMX-U\end{tabular}}}        & 1865               & 12 hr                & 500$\mu$m              & 0.2ms                  \\ \hline
\end{tabular}
\caption{Parallelization and Solver Times for the Shercliff, Dam Breaking, and LMX-U Cases}
\label{tab:cpuhours}
\end{table}

The Pressure-Velocity Coupling method uses the Pressure Implicit with Splitting of Operations (PISO) algorithm, for coupling the continuity and momentum equations. 
For both solid and fluid regions, the electrical potential equation is solved. 
Additionally, for fluid regions, there is an interface-capturing procedure, then the density is updated, the velocity flux is calculated for each phase, and then the current density is calculated. 
For both solid and fluid regions, the boundary conditions can be updated and looped again or advance to the next time step. 
The PISO loop is completed to obtain the corrected velocity and is combined with additional iterations that include the solid regions, so the solution process becomes similar to a combination of PISO and SIMPLE algorithms. 
The fluid and solid regions are coupled by solving each one separately and then exchanging information through boundary conditions that contain details from the neighboring region. This requires outer loops to ensure proper coupling within each time step. 

For interface capturing, the MUlti dimensional Limiter for Explicit Solution (MULES) method modifies the advection of the volume fraction by adding an interface compression velocity term. This controls the thickness and reduces smearing of the interface.

For all cases here involving a magnetic field, the turbulence can be considered suppressed. This is characterized by the critical value of the ratio of the Reynolds number and Hartmann number ($R=Re/Ha$), which was experimentally found to be approximately $R_{crit} = 380$ through a transition to turbulence in the Hartmann layer electromagnetic flow in a square duct \cite{moresco2004experimental}. This ratio can be interpreted as the Reynolds number based on the thickness of the Hartmann layer and indicates that the magnetic field can suppress turbulence for cases of $R$ lower than the critical value. 

For cases with no magnetic fields (Dam Breaking and LMX with No Magnetic Field), the value of $R > R_{crit}$. Therefore, here a turbulence model is used, since in this regime, turbulence cannot be assumed to be suppressed and needs to be modeled. The standard k-$\epsilon$ model was chosen due to its reliability and relative ease of use, and common usage in CFD simulations, where it has been used successfully in open channel flows \cite{farhadi2018accuracy,shaheed2019comparison}. 
This involved only slight changes to the case setup, involving the initial and boundary conditions of the turbulence variables, as well as the inclusion of the k-$\epsilon$ turbulence equations. 
Conversely, for all cases with magnetic fields, the setup does not include any turbulence equations, and therefore no MHD turbulence is considered.

\subsection{Conditions for Numerical Stability and Accuracy}

The Courant number describes how fast a fluid travels across a grid of cell length $\Delta x$ in time $\Delta t$, with a characteristic velocity $U$, and is generally defined as  
$Co = U \Delta t / \Delta x $. 
Extending to a discretized form, the volumetric flux is first evaluated using the scalar product of velocity $\mathbf{U}_f$ and the surface normal vector $\mathbf{S}_f$ of face $f$, and then summed for all $N_f$ faces. Then the time step $\Delta t$ and cell volume $V_C$ of a cell $C$ are used to calculate  
$Co_C = \frac{\Delta t}{2 V_C} \Sigma_{f=1}^{N_f} U_f \cdot S_f $ 
for all cells. 
The Courant–Friedrichs–Lewy (CFL) condition states that $Co \leq Co_{C,max}$ where $Co_{C,max}$ is typically less than or equal to one. 
With a stationary grid, this is implemented by limiting the time step such that the maximum courant number is not exceeded.

The Alfvén Velocity ($v_A=B/\sqrt{\mu_0 \rho}$) is the speed at which Alfvén waves propagate and indicates how quickly disturbances in the magnetic field are communicated through the fluid. This can be useful in predicting and controlling the stability of liquid metal flows, and characteristic values for Galinstan, NaK, and Lithium are listed in Table \ref{tab:times}. The magnetic damping time is calculated from values of the fluid density, electrical conductivity, and a characteristic magnetic field, $\tau_B = \rho/(\sigma B^2)$,  and represents the time scale on how quickly the magnetic field can dissipate the kinetic energy of the flow. The solver time step $\Delta t$ should be less than $\tau_B$, with typical values in Table \ref{tab:times}.

\begin{table}[h!]
\centering
\small 
\begin{tabular}{|>{\centering\arraybackslash}m{2cm}|>{\centering\arraybackslash}m{2cm}|>{\centering\arraybackslash}m{2cm}|>{\centering\arraybackslash}m{2cm}|}
\hline
\textbf{Material} & \textbf{Alfvén Velocity, \( v_A \) (m/s)} & \textbf{Alfvén Time, \( \tau_A \) (s)} & \textbf{Magnetic Damping Time, \( \tau_B \) (s)} \\
\hline
\textbf{Galinstan} & 151.35 & $3.30 \times 10^{-4}$ & $2.28 \times 10^{-2}$ \\
\hline
\textbf{NaK} & 370.57 & $1.46 \times 10^{-4}$ & $3.22 \times 10^{-4}$ \\
\hline
\textbf{Lithium} & 3970.02 & $2.52 \times 10^{-5}$ & $5.89 \times 10^{-6}$ \\
\hline
\end{tabular}
\caption{Alfvén velocity, Alfvén time, and magnetic damping time for Galinstan, NaK, and Lithium.}
\label{tab:times}
\end{table}


\subsection{Cases Tested}



There are a variety of cases with various conditions, and non-dimensional parameters are useful for the characterization of the regime of the various flows. 
The Reynolds number ($\mathrm{Re=UL/\nu}$) represents the ratio of inertial and viscous forces. 
The Hartmann number ($\mathrm{Ha=BL\sqrt{\sigma/(\rho\nu)}}$) represents the relative effect of the magnetic field, where  Ha$^2$ is the ratio of magnetic and viscous forces. 
Here, L, B, and U are characteristic values of length, magnetic field, and velocity on a per-case basis. The characteristic length for the Hartmann number is typically half of the distance across the direction parallel to the magnetic field. For example, the length could be the half-width of a rectangular duct or the radius of a circular pipe. The interaction parameter is the ratio of electromagnetic to inertial forces and can be calculated as $\mathrm{N=Ha^2/Re}$. 
The Magnetic Bond number ($\mathrm{Bo_m}$) is the ratio of magnetic Lorentz forces to surface tension forces, which is typically expressed as  $(B^2 \ell) / (\mu_0 \sigma_{st})$ but here will be taken as ($JB$)/($\sigma_{st}/\ell^2)$, to consider the current density in general. 
Lastly the wall conductance ratio, c = $(\sigma_w t_w)/(\sigma \ell)$, where $\sigma_w $ and $t_w$ are the electrical conductivity and thickness of the solid wall.

\begin{figure}
    \centering

    \begin{subfigure}[t]{0.45\columnwidth}
        \centering
        \includegraphics[width=\textwidth]{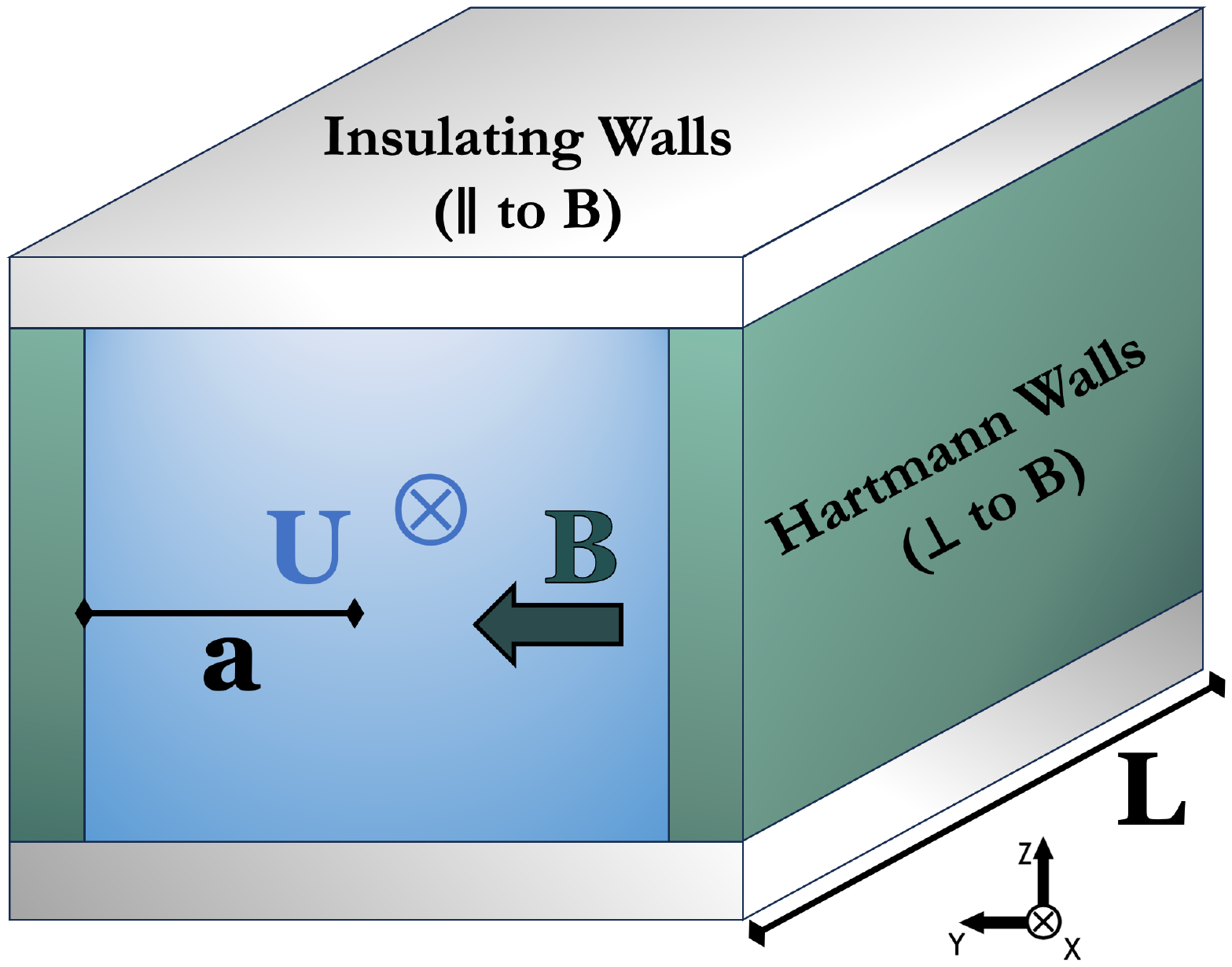}
        \caption{Closed Channel for Shercliff \cite{shercliff1953steady} and Hunt \cite{hunt1965magnetohydrodynamic}. Walls perpendicular and parallel to the magnetic field vary in electrical conductivity}
        \label{fig:ClosedChannelDiagram}
    \end{subfigure}
    \hfill
    \begin{subfigure}[t]{0.45\columnwidth} 
        \centering
        \includegraphics[width=\textwidth]{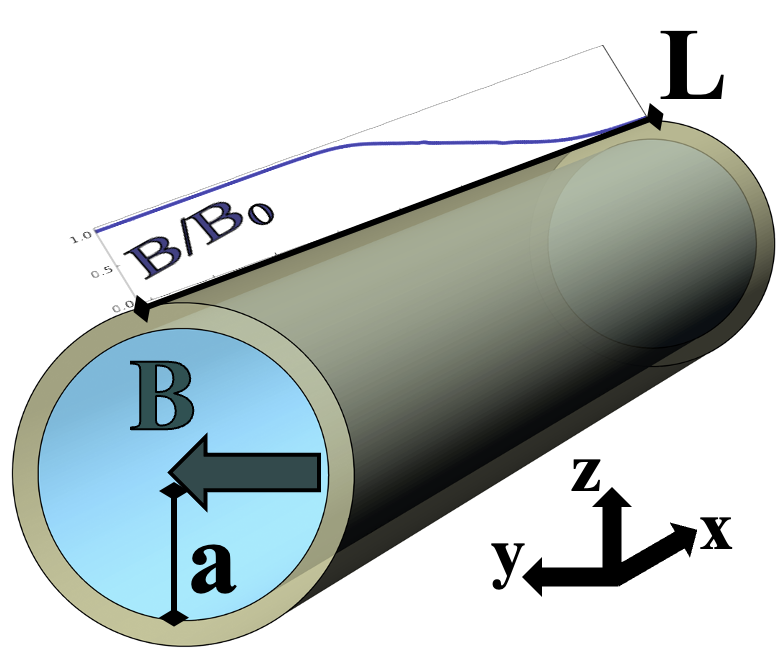}
        \caption{Pipe in Fringing Magnetic Field \cite{buhler2020experimental}. Flow exits magnetic field}
        \label{fig:Buhlerpipe_setup}
    \end{subfigure}

    \medskip

    \begin{subfigure}[t]{0.45\columnwidth}
        \centering
        \includegraphics[width=\textwidth]{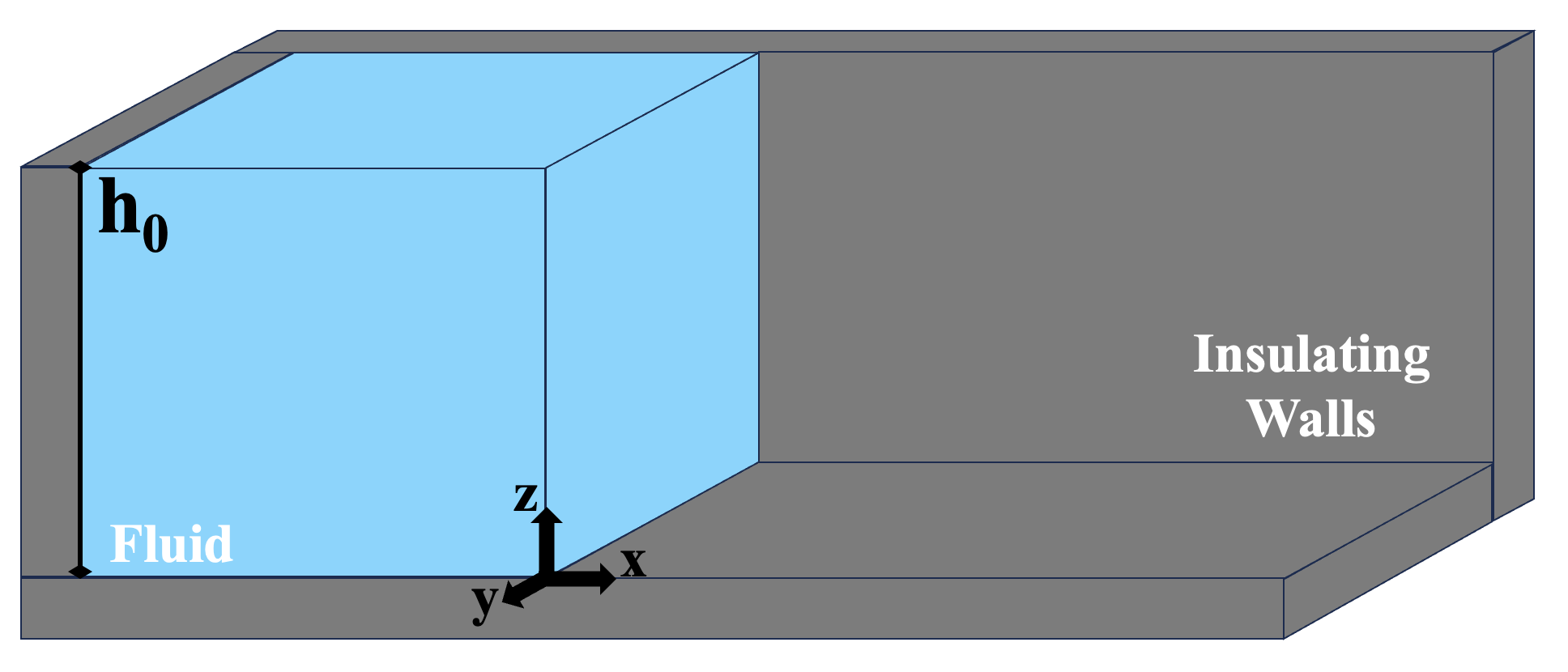}
        \caption{Dam Breaking Case\cite{ozmen2010dam}, Fluid is Released from Rest with Initial Height of h$_0$}
        \label{fig:damBreak_setup}
    \end{subfigure}
    \hfill
    \begin{subfigure}[t]{0.45\columnwidth}
        \centering
        \includegraphics[width=\textwidth]{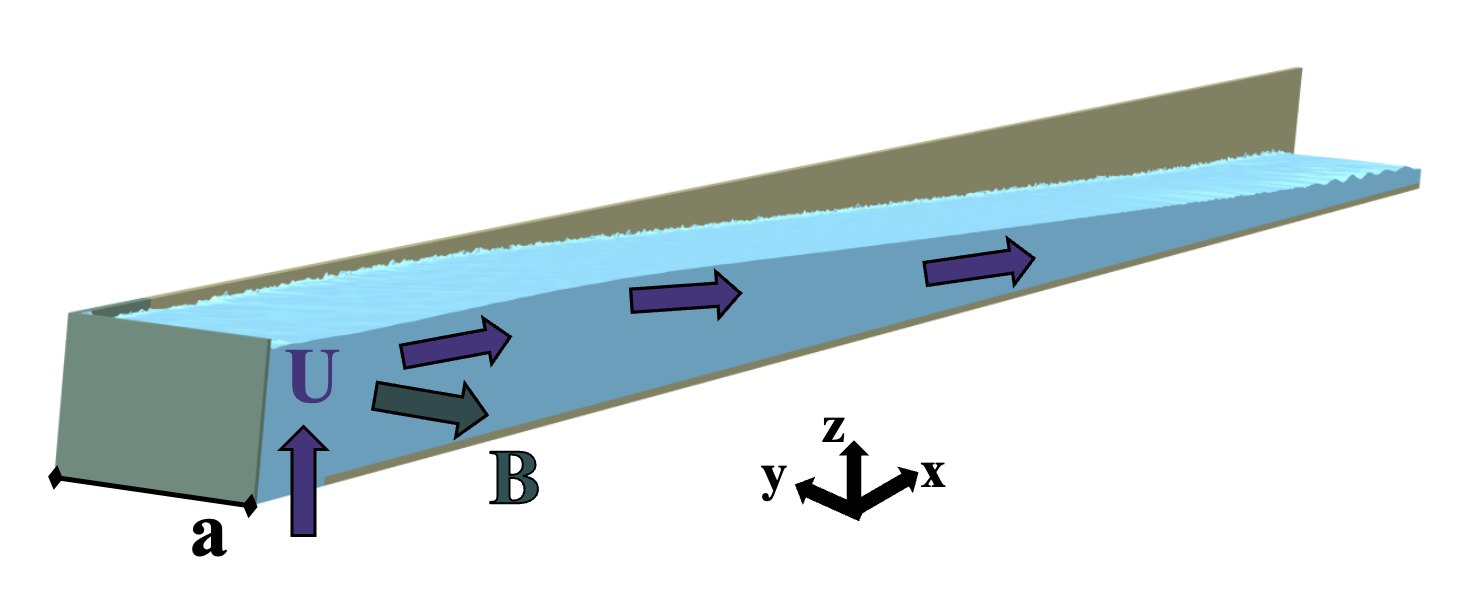}
        \caption{LMX-U Setup \cite{sun2023magnetohydrodynamics}. Flow along free surface channel across transverse magnetic field}
        \label{fig:lmx_setup}
    \end{subfigure}

        \medskip

    \begin{subfigure}[t]{0.45\columnwidth}
        \centering
        \includegraphics[width=\textwidth]{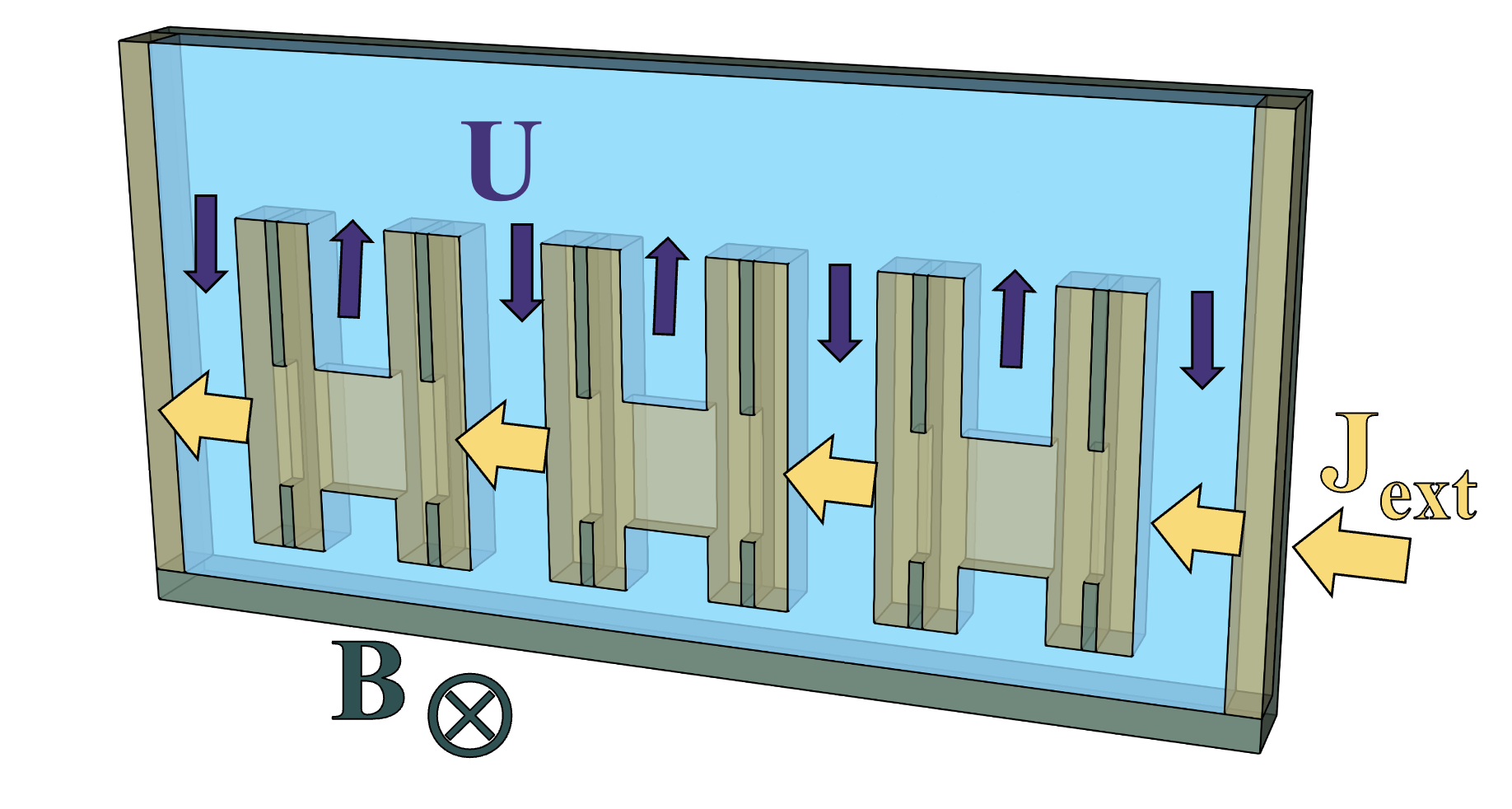}
        \caption{Divertorlets Setup \cite{saenz2022divertorlets}. Applied current with external magnetic field induces opposing flows in opposite channels}
        \label{fig:divertorlets_setup}
    \end{subfigure}
    \hfill

    \caption{All Case Setups}
    \label{fig:AllCaseSetups}
\end{figure}

Two analytical models will be considered to provide numerical verification, involving liquid metal flows in square ducts across transverse magnetic fields. 
First, the Shercliff Case is a closed channel duct with all insulating walls \cite{shercliff1953steady}. 
Second, the Hunt Case is also with a closed channel duct, but by contrast to the Shercliff Case has conducting walls perpendicular to the magnetic field and insulating walls parallel to the magnetic field \cite{hunt1965magnetohydrodynamic}. Figure \ref{fig:ClosedChannelDiagram} shows the setup for the Shercliff and Hunt Cases, which have a wall conductance ratio of $\mathrm{c}=0$ and $0.05$ respectively. 
In both cases, the presence of the magnetic field alters the boundary layer structure, reducing the boundary layer thickness -- i.e., the distance to where the flow reaches its bulk flow velocity. 
The Hartmann layer thickness scales as 1/Ha, so this boundary layer becomes very thin with increasing magnetic field magnitude, requiring higher density mesh close to the wall to be resolved. To ensure the changes in velocity and electric potential are captured, at least 10 cells are included in the boundary layer.  

Next, a variety of cases were chosen for experimental validation. These cases test the ability of FreeMHD in various conditions such as fringing fields, free surface evolution with no magnetic fields, and then combined with fringing fields and free surface flows. 

The first case experimental validation, the fringing magnetic field, involves sodium potassium (NaK) flow through a dipole magnet in a closed pipe with electrically conducting walls \cite{buhler2020experimental}. The experimental measurements focus on the region exiting the magnet as shown in Figure \ref{fig:Buhlerpipe_setup}. This case involves 3D currents and is calculated on a non-orthogonal grid. 


The next experimental validation case involves a dam breaking case setup, with water in a channel initially at rest and then released, with the free surface profile being measured \cite{ozmen2010dam}. The case setup is shown in Figure \ref{fig:damBreak_setup}, where the initial height is $h_0$ which is used for scaling the results. The characteristic velocity can be taken as $\sqrt{g h_0}$, with a characteristic length of $h_0$ to give a Reynolds number of 3.9E5. 

Another case for comparison is the Liquid Metal eXperiment Upgrade (LMX-U), which is a test bed for liquid metal experiments with a flow loop and free surface channel \cite{sun2023magnetohydrodynamics}. 
The LMX-U results are used for free-surface liquid metal flow experimental validation using galinstan, where Figure \ref{fig:lmx_setup} shows the setup.
In this case, the inlet is vertical, and levels off downstream, with a channel length from inlet to outlet of L = 1.2m. 
Another characteristic length for an open channel is the hydraulic radius ($\mathrm{r_H=A/P})$, the ratio of the cross-sectional area of the fluid (A=hw) divided by the wetted perimeter (P=2h+w).
For LMX-U, with an average height of 2 cm and width of 10 cm, the hydraulic radius is 1.4 cm. 
With flow rate (Q) in the experiment at 16.3 L/min, the average velocity is Q/A = 0.1358 m/s. Therefore, using the hydraulic radius and average velocity as the characteristic length and velocity, the Reynolds Number is 6,380. 
The main characteristic length is the channel half-width, $\ell$ = 0.05 m, which is used for non-dimensionalization of the results: the x-axis for the flow direction as well as the fluid height. This length is also used when calculating the Hartmann number. 
The galinstan flows across the transverse magnetic field, where the magnet is located from x/$\ell$ = 0 to 14.8. The test cases of magnetic field strengths of 0.3, 0.2, and 0 T therefore correspond to Hartmann numbers of 607, 404, and 0. 
For the Ha = 0 case, a turbulence model was used (k-$\epsilon$), as the turbulence was not suppressed. 
The wall conductance ratio with a copper liner is approximately 0.8. 
For the magnetic bond number, taking $\sigma U B$ as a representative scale for $J$, the 0.2 and 0.3 T cases correspond to ($\sigma U B^2$)/($\sigma_{st}/\ell^2)$ = 79 and 180. 

Divertorlets is the last case for experimental validation, and represents an alternative LM-PFC design, with applied currents inducing flows between slats \cite{saenz2022divertorlets}. As with LMX-U, this device has been tested at PPPL using galinstan, and the Divertorlets results are used for further free-surface liquid metal flow experimental validation, with Figure \ref{fig:divertorlets_setup} showing the setup. A characteristic length and velocity can be taken as the distance between channel slats (5.1 mm) and the average vertical velocity between channels (0.25 m/s), to give a Reynolds number of 4,279. The half-width between channel slates (2.55 mm) and the applied magnetic field applied (0.2 T) is used for a Hartmann Number of 21. 

The applicable Non-Dimensional Parameters for each case are summarized in Table \ref{tab:NonDimensionalParameters}: the Reynolds Number, Hartmann Number, Interaction Parameter, Magnetic Reynolds Number, and Magnetic Bond Number.

\begin{table}
    \centering
    \begin{tabular}{l|ccccc} 
        \hline
         \textbf{Case Name}&  $\mathbf{Re}$&  $\mathbf{Ha}$&  $\mathbf{N}$&  $\mathbf{Re_m}$& $\mathbf{Bo_m}$\\ 
         \hline
         Closed Channel Duct&  1.0E2&  1.0E3&   1.0E4&  1.2E-1& ---\\ 
         Pipe Fringing B Field&  2.0E4&  2.0E3&  2.0E2&  6.4E-2& ---\\ 
         Dam Breaking&  3.9E5&  ---&  ---&  ---& ---\\ 
         LMX-U&  6.4E3&  6.1E2&  5.8E1&  2.6E-2& 1.8E2\\ 
         Divertorlets&  4.3E3&  2.1E1&  1E-1&  5.0E-3& 1.5E0\\ 
         \hline
    \end{tabular}
    \caption{Characteristic Non-Dimensional Parameters}
    \label{tab:NonDimensionalParameters}
\end{table}

\subsection{Mesh Generation}
\begin{figure}
    \centering

    \begin{subfigure}[t]{0.45\columnwidth}
        \centering 
        \includegraphics[width=\textwidth]{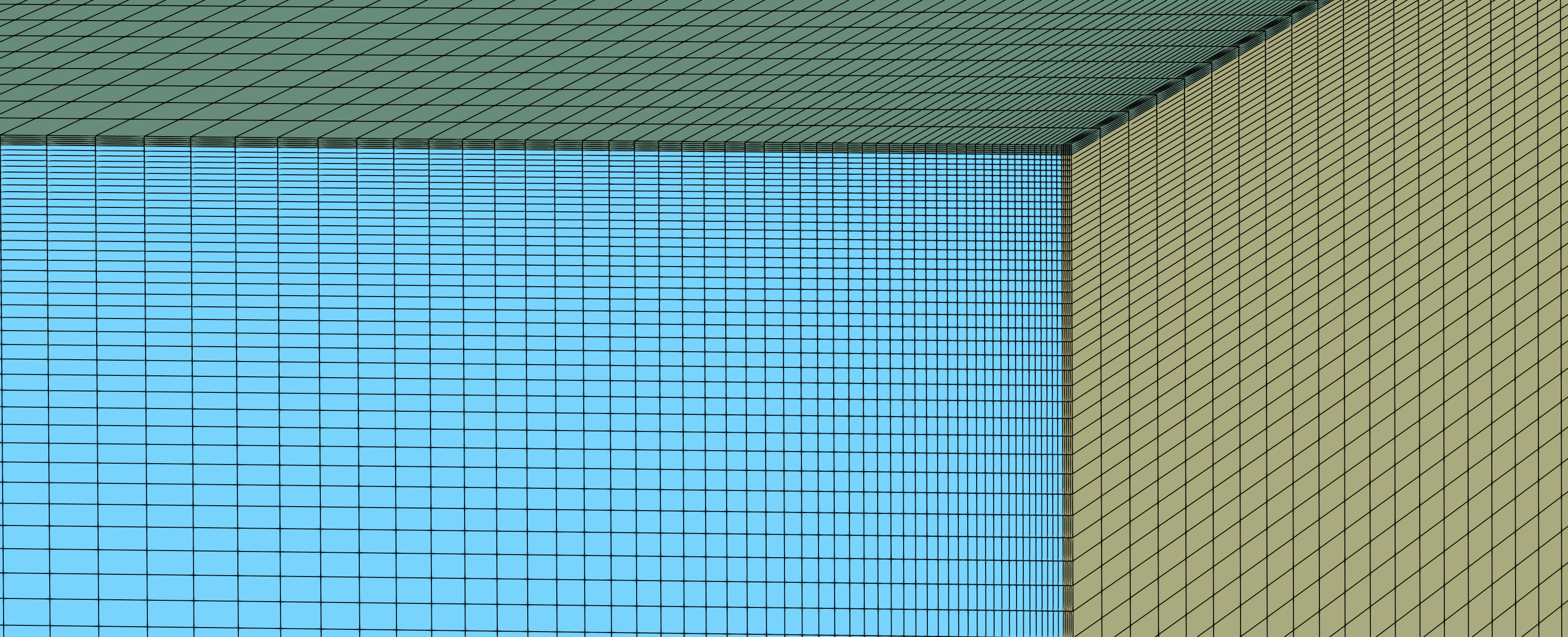}
        \caption{Shercliff Mesh}
        \label{fig:shercliff_mesh_Ha100}
    \end{subfigure}
    \hfill
    \begin{subfigure}[t]{0.45\columnwidth}
        \centering
        \includegraphics[width=\textwidth]{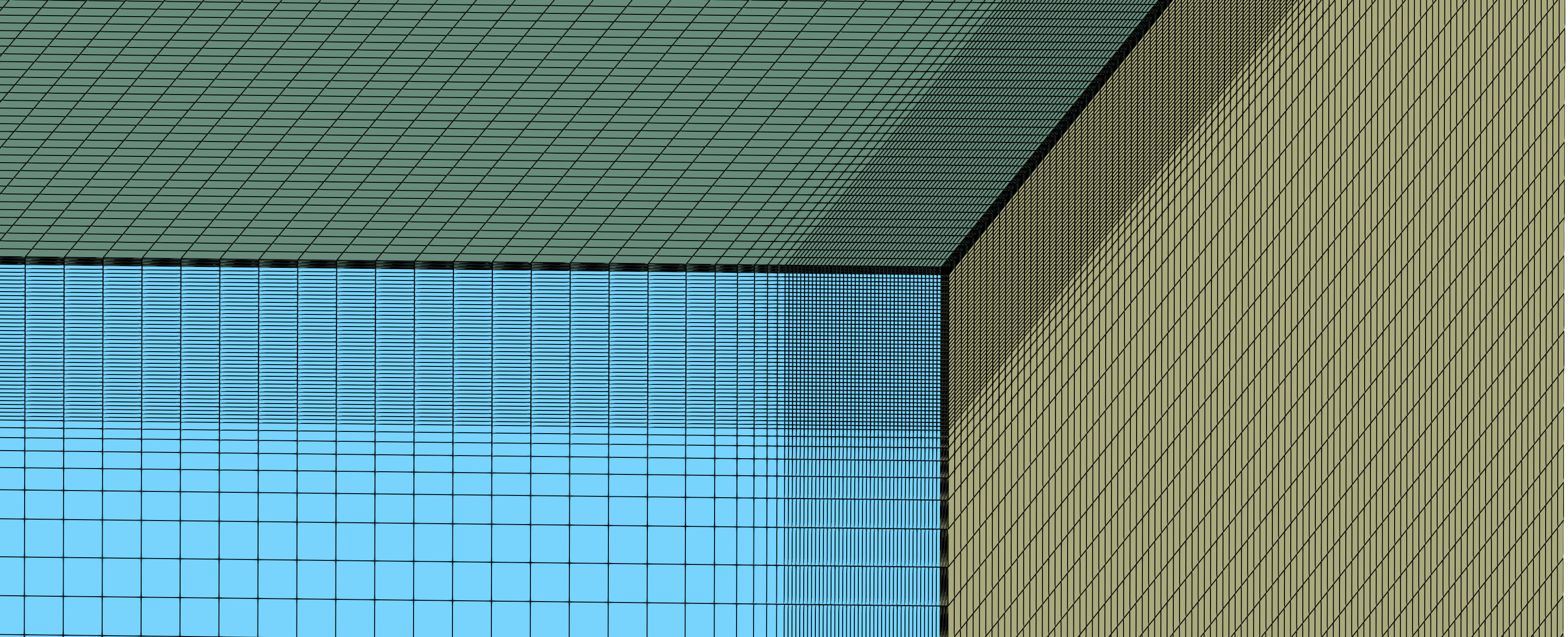}
        \caption{Section of Hunt Mesh, Boundary Layer}
        \label{fig:hunt_mesh_Ha1000}
    \end{subfigure}

    \medskip

    \begin{subfigure}[t]{0.45\columnwidth}
        \centering 
        \includegraphics[width=\textwidth]{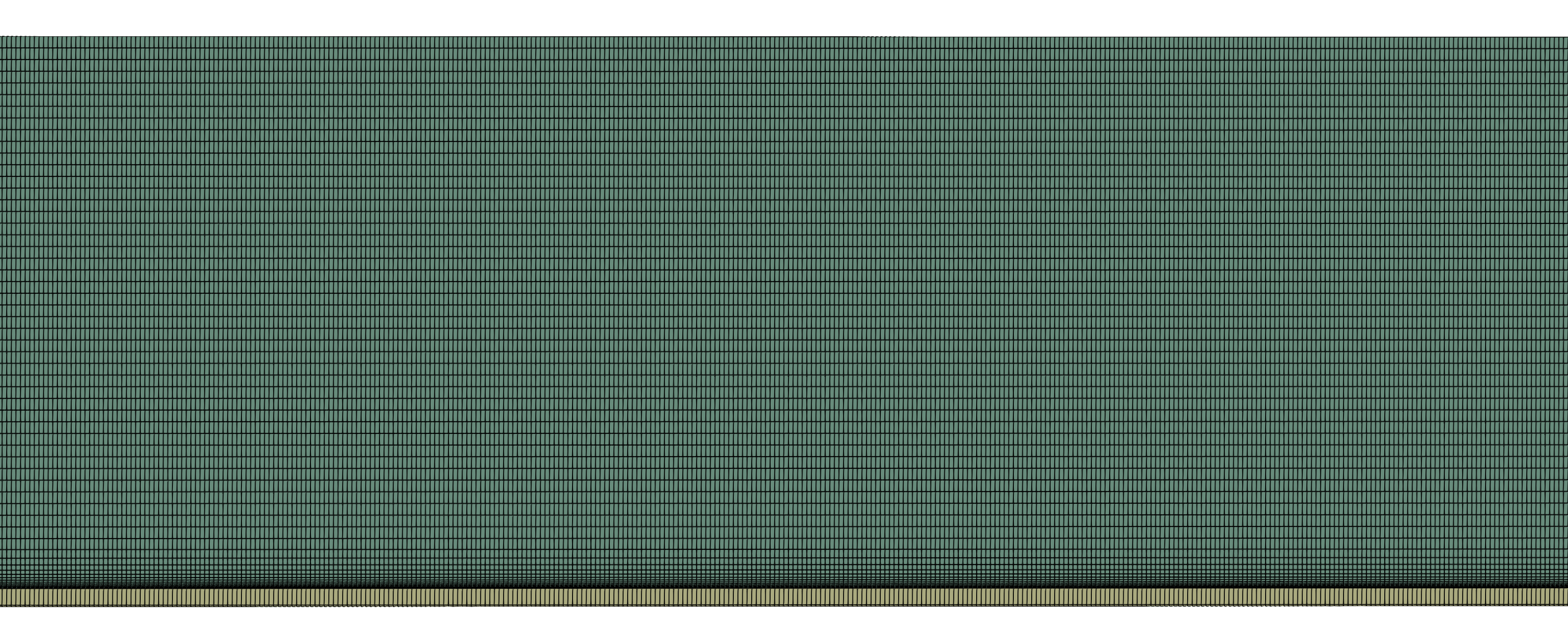}
        \caption{Dam Breaking Mesh}
        \label{fig:dam_mesh}
    \end{subfigure}
    \hfill
    \begin{subfigure}[t]{0.45\columnwidth}
        \centering
        \includegraphics[width=\textwidth]{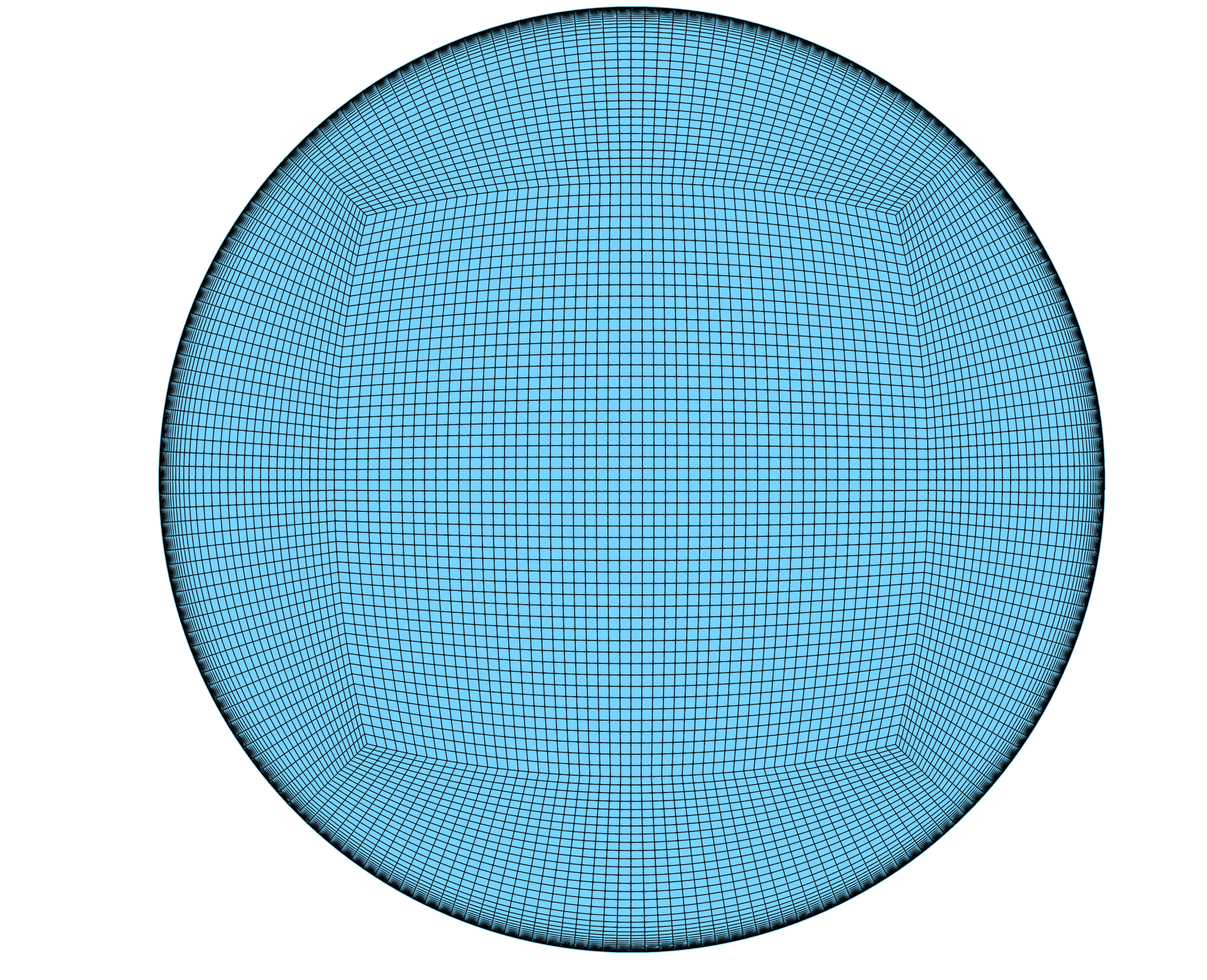}
        \caption{Pipe Mesh Case}
        \label{fig:pipe_mesh}
    \end{subfigure}

    \medskip

    \begin{subfigure}[t]{0.45\columnwidth}
        \centering
        \includegraphics[width=\textwidth]{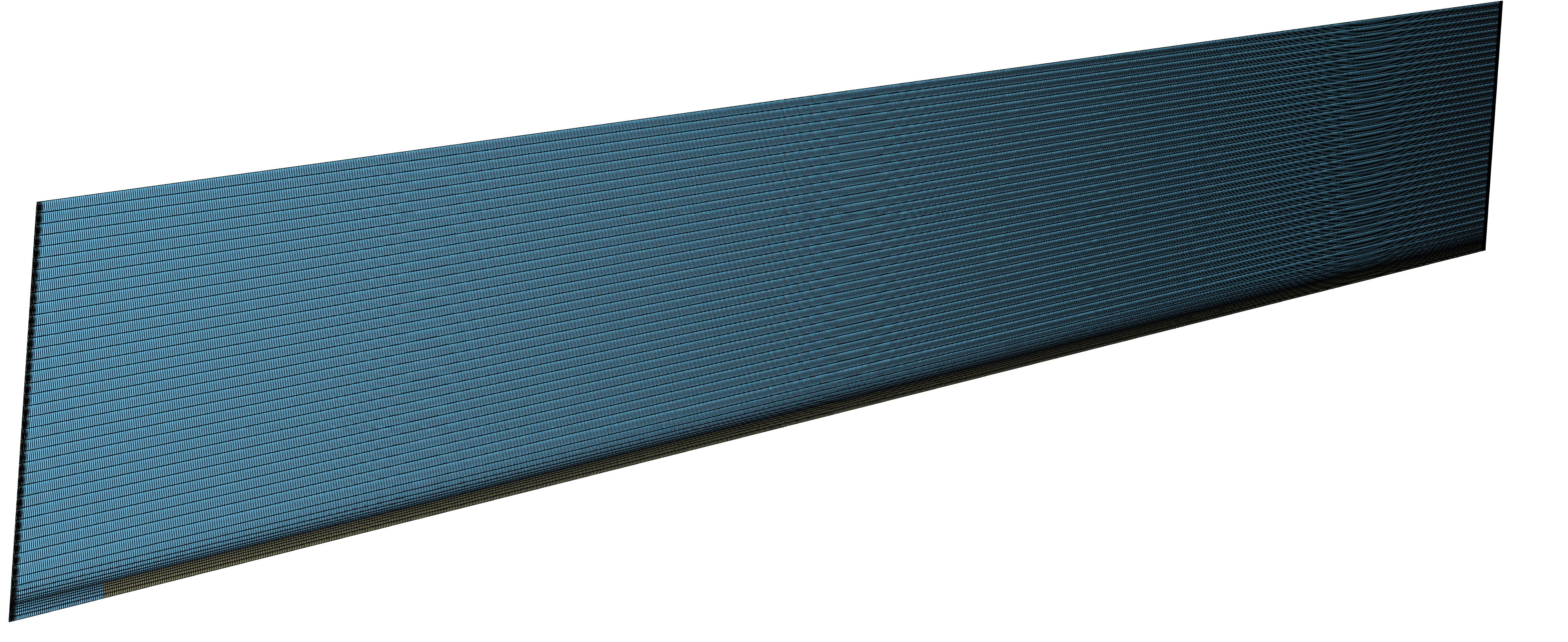}
        \caption{LMX-U Mesh}
        \label{fig:lmx_mesh}
    \end{subfigure}
    \hfill
    \begin{subfigure}[t]{0.45\columnwidth}
        \centering
        \includegraphics[width=\textwidth]{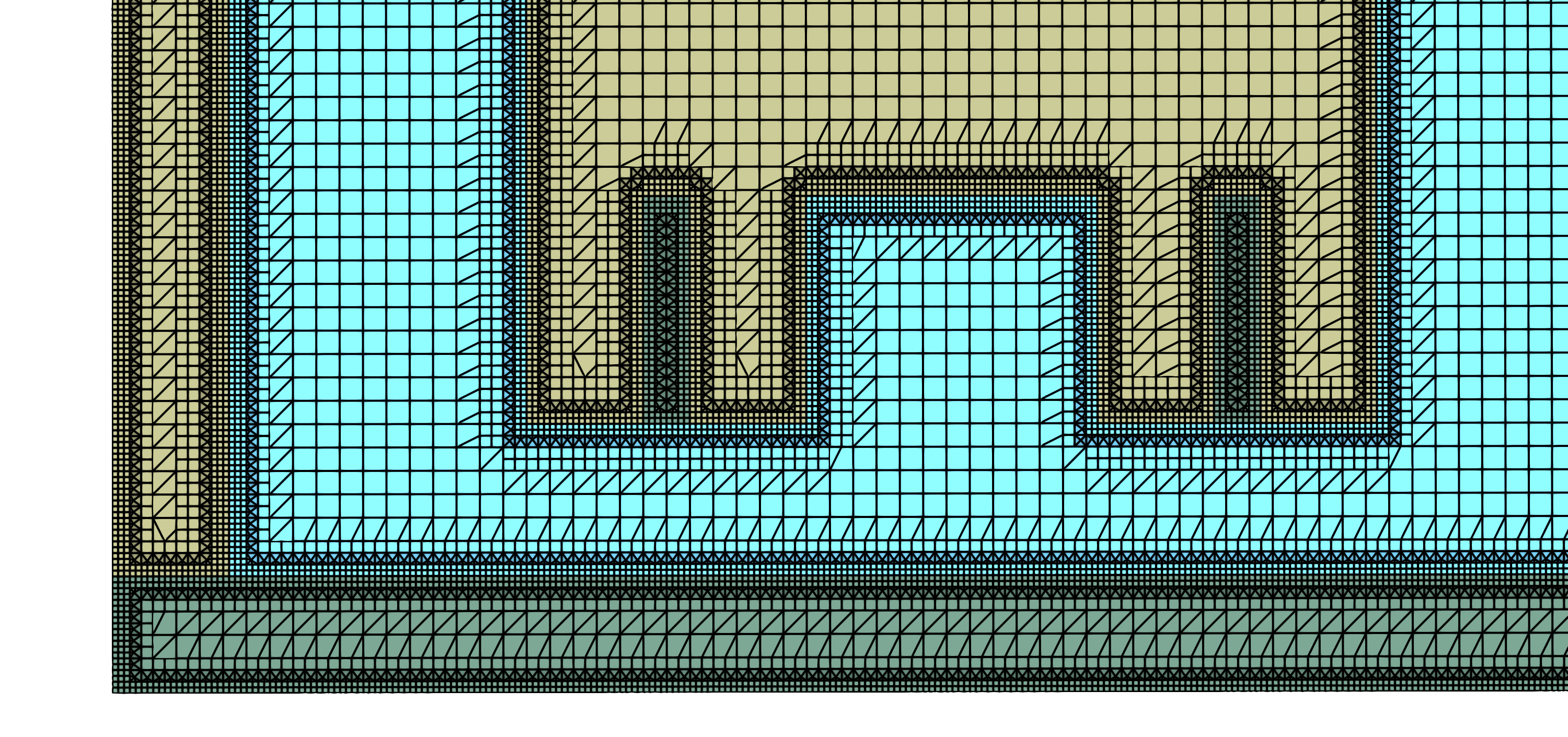}
        \caption{Section of Divertorlets Mesh}
        \label{fig:divertorlets_mesh}
    \end{subfigure}

    \caption{Examples of Mesh}
    \label{fig:AllMeshes}
\end{figure}

The built-in OpenFOAM \texttt{blockMesh} was used for rectangular channels with various regions for the closed channel cases. Cylindrical mesh generation required special consideration, with a $M4$ script to specify meshes for the fringing magnetic field in a pipe case. Another option used is \texttt{snappyHexMesh} which is useful for boundary layers on geometries that do not form straightforward boxes or cylinders. A final meshing generation option has been done with Salome, an open-source software, which is converted to OpenFOAM meshes using the \texttt{ideasUnvToFoam} conversion. Examples of the meshes for the cases tested are shown in Figure \ref{fig:AllMeshes}.

\section{Results}
\label{section:Results}
\subsection{Verification with Analytical Solutions: Shercliff and Hunt Flow}
\label{subsection:shercliffhunt}
To verify the model, FreeMHD can be compared to theoretical results for which there are analytical solutions, and using a reformulation of Hunt’s equation for velocity profiles of fully developed MHD flows, for high Hartmann numbers \cite{ni2007current}. FreeMHD is verified using analytical solutions for closed channel flows by examining the effect on the closed channel velocity profile and pressure drop while varying the Hartmann number and wall conductance ratio. 
The two cases presented here are validated using 3D geometry, with channels long enough for the flow to become fully developed. 

\begin{figure}
    \centering

        \begin{subfigure}{\linewidth}
        \centering
        \includegraphics[width=\linewidth]{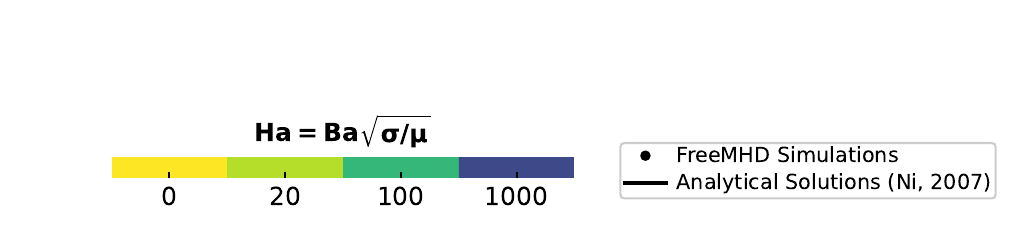}
    \end{subfigure}

    \hfill

    \begin{subfigure}[t]{0.45\columnwidth}
        \centering
        \includegraphics[width=\linewidth]{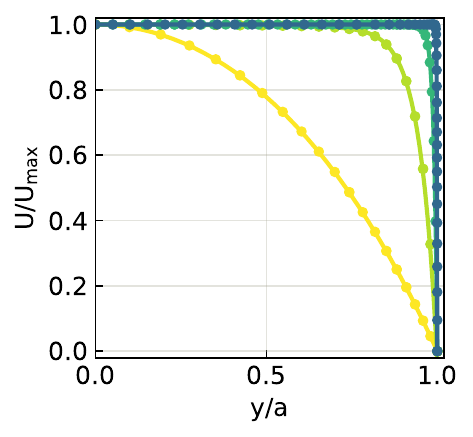}
        \caption{Shercliff Hartmann Layer}
        \label{fig:ShercliffHartmannLayer}
    \end{subfigure}
    \hfill
    \begin{subfigure}[t]{0.45\columnwidth}
        \centering
        \includegraphics[width=\textwidth]{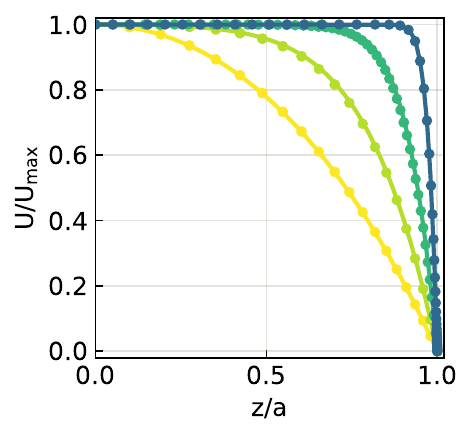}
        \caption{Shercliff Side Layer}
        \label{fig:ShercliffSideLayer}
    \end{subfigure}

    \hfill

    \begin{subfigure}[t]{0.45\columnwidth}
        \centering
        \includegraphics[width=\textwidth]{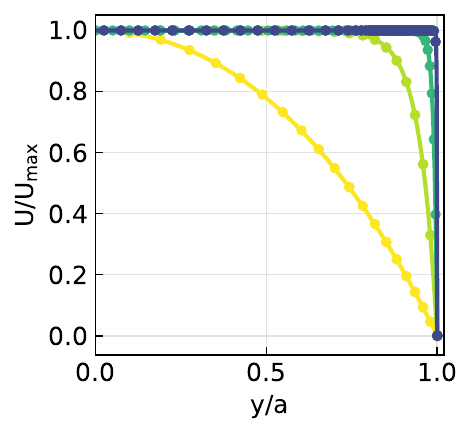}
        \caption{Hunt Hartmann Layer}
        \label{fig:HuntHartmannLayer}
    \end{subfigure}
    \hfill
    \begin{subfigure}[t]{0.45\columnwidth}
        \centering
        \includegraphics[width=\textwidth]{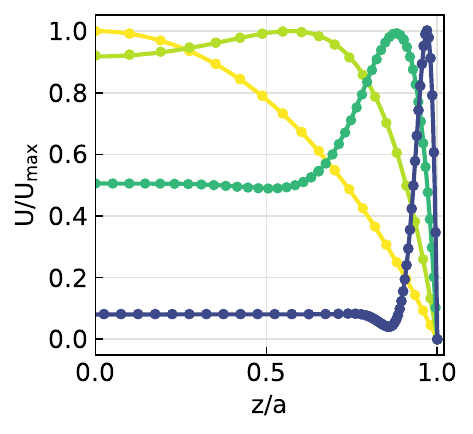}
        \caption{Hunt Side Layer}
        \label{fig:HuntSideLayer}
    \end{subfigure}
    
    \caption{Comparison to Closed Channel Results of Fully Developed Velocity Profiles,  Shercliff Insulating Walls (c=0) and Hunt Conducting Walls (c=0.05), Hartmann ($\perp$B) and Side ($\parallel$B) Layers. Points are FreeMHD results from the end of the channel (x=L), lines are analytical solutions \cite{ni2007current}}
    \label{fig:ClosedChannelResults}
\end{figure}

The validation results for the Shercliff case are shown in Figure \ref{fig:ShercliffHartmannLayer} and \ref{fig:ShercliffSideLayer} for the Hartmann Layer and Side Layer, plotting the normalized velocity profile across the channel half-width, where the points are FreeMHD simulation results and the lines are analytical solutions using the reformulation for high Hartmann numbers \cite{ni2007current}.
The non-dimensionalized flow rate for the Shercliff case is compared with solutions in Table \ref{tab:Shercliff}.  

Likewise, the validation results for the Hunt case (conductance ratio, $c = 0.05$) are shown in Figure \ref{fig:HuntHartmannLayer} and \ref{fig:HuntSideLayer} for the Hartmann Layer and Side Layer, again plotting the normalized velocity profile across the channel half-width.

\begin{table}[]
    \small
    \centering
\begin{tabular}{|c|c|c|c|}
\hline
\multicolumn{1}{|l|}{Ha} & \multicolumn{1}{l|}{Q$_\mathrm{FreeMHD}$} & \multicolumn{1}{l|}{Q$_\mathrm{Solution}$} & \multicolumn{1}{l|}{Percent Error (\%)} \\ \hline
0                        & 5.653E-1                            & 5.623E-1                       & 0.534                              \\ \hline
20                       & 1.604E-1                            & 1.567E-1                       & 2.36                               \\ \hline
100                      & 3.762E-2                            & 3.621E-2                       & 3.89                               \\ \hline
1000                     & 3.899E-5                            & 3.887E-5                       & 0.309                              \\ \hline
\end{tabular}
\caption{Non-dimensionalized flow rates compared to the analytical solution of the Ha = 0 Case\cite{shah1980laminar} and Ha $>$ 0 Shercliff Cases Case\cite{ni2007current}}
\label{tab:Shercliff}
\end{table}

\subsection{Validation with Experiments: Fringing field closed pipe}

\begin{figure}
    \centering
    \includegraphics[width=\linewidth]{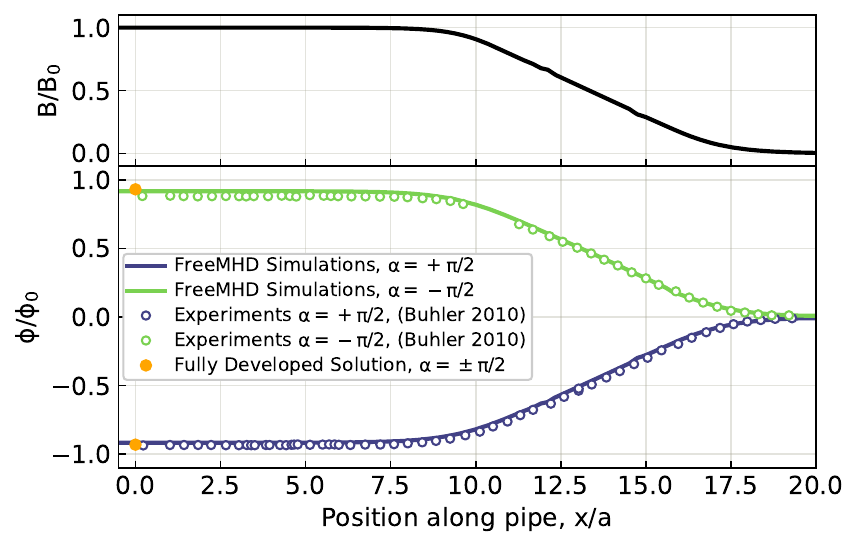}
    \caption{Magnetic Field Distribution entering pipe and Non-dimensional electric potential at the wall at both sides along the pipe, where x=0 is the magnet center and a is the pipe radius\cite{buhler2020experimental}}
    \label{fig:Buhler2020_combinedBAndElectricPotential}
\end{figure}

Next, FreeMHD is compared for validation to the experimental results of pipe flow through a fringing magnetic field \cite{buhler2020experimental}. 
The fringing magnetic field leads to 3D pressure and current density distributions that are distinct from the fully developed solutions. 
The electric potential ($\phi$) was measured at the walls of the pipe, and has been scaled by $\phi_0=u_0LB_0$. The non-dimensional electric wall potential is compared here, in Figure \ref{fig:Buhler2020_combinedBAndElectricPotential} with $\alpha=\pm \frac{\pi}{2}$ corresponding to the left and right sides of the pipe parallel to the magnetic field.  Additionally, the fully developed solution is plotted, where the experimental and simulation results at x=0 in the center of the magnet demonstrate relative agreement with the fully developed solution in this region. 

\subsection{Validation with Experiments: Dam Break Experiment (Free Surface, No MHD) \cite{ozmen2010dam}}

The validation of the free surface was done through comparison with a dam breaking experiment \cite{ozmen2010dam}, where the evolution of the free surface can be used to validate the application of the VoF method. 

\begin{figure}
    \centering
    \includegraphics[width=\linewidth]{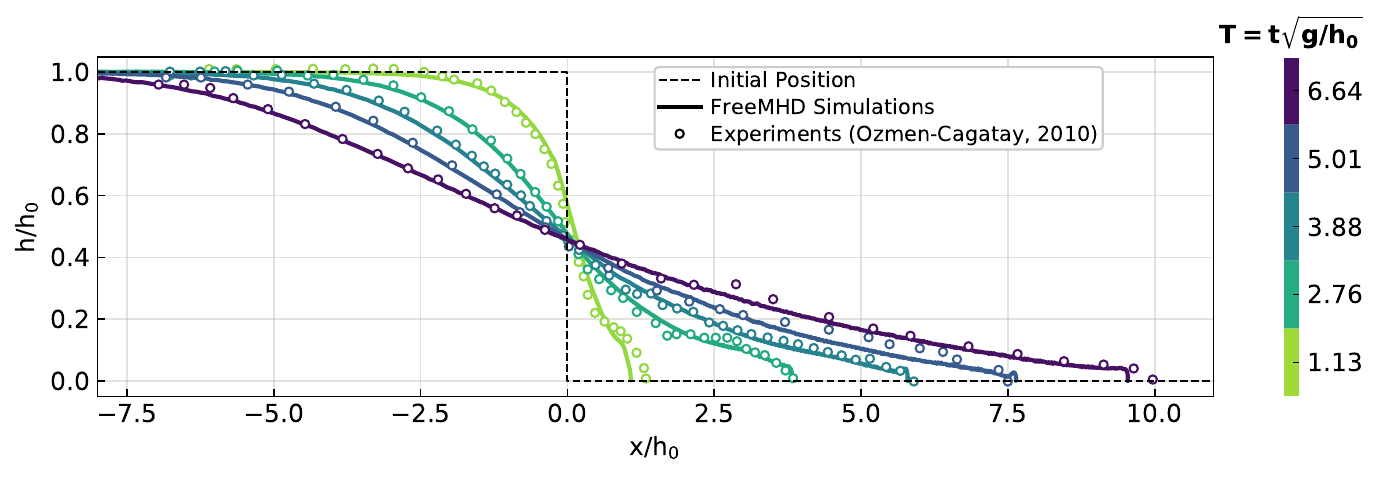}
    \caption{Evolution of the free surface of a dam breaking, where the height and position are scaled by $\mathrm{h_0=0.25 m}$ \cite{ozmen2010dam} }
    \label{fig:DamBreakOzmenCagatay2010}
\end{figure}

Here, a rectangular channel of 9 m length, 0.3 m width, and 0.34 m height was used with an initial reservoir height of $h_0=0.25$ m.  The free surface profiles were measured optically at 50 frames/s. FreeMHD was used to calculate using a 3D domain, and the results were normalized by $h_0$ and plotted in Figure \ref{fig:DamBreakOzmenCagatay2010}. As can be seen, there is very good agreement between the experimental results and FreeMHD simulations, with root mean square error of the profiles between experiment and simulation on the order of 2\%.

\subsection{Validation with Experiments: Free Surface MHD Experiments (LMX-U)}
\label{subsection:FreeSurfaceMHD}

\begin{figure}
    \centering
    \includegraphics[width=\linewidth]{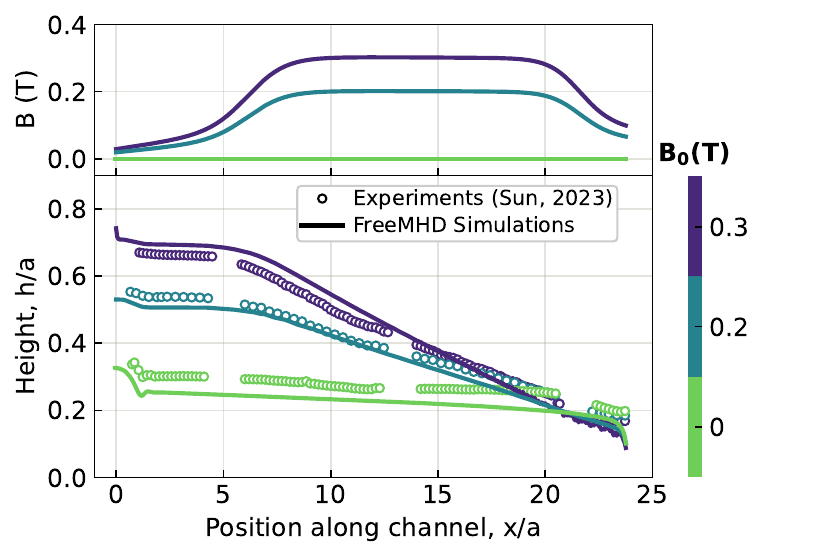}
    \caption{LMX-U Experimental measurements and FreeMHD simulation results, comparison of liquid metal profiles along the length of the channel, with depths and positions normalized by channel half-width, with magnetic field distributions used in FreeMHD simulations.} 
    \label{fig:LMX_combinedBAndHeight}
\end{figure}

Next, combining MHD and free surface flows for a validation case on LMX-U, a flow loop of galinstan in an open channel across an electromagnet at PPPL. \cite{sun2023magnetohydrodynamics}. 
The magnetic field profile is shown in the top of Figure \ref{fig:LMX_combinedBAndHeight}, where the flow is across the magnet from left to right. Experiments measured the height of liquid metal using a laser-camera system. The height profile of the free surface is shown, and the simulations are plotted at the centerline. The flow is across magnetic fields of 0, 0.2, and 0.3 T.

\subsection{Validation with Experiments: Free Surface MHD Experiments (Divertorlets)}
\begin{figure}
    \centering
    \includegraphics[width=\linewidth]{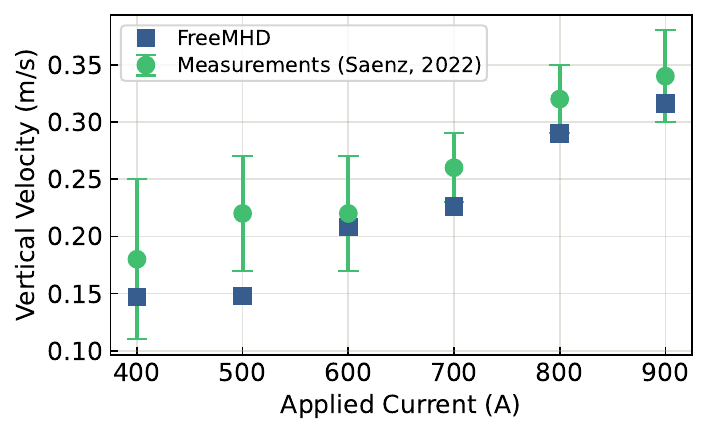}
    \caption{Divertorlets Average Velocity Between Slats \cite{saenz2022divertorlets}}
    \label{fig:DivertorletsVerticalVelocity}
\end{figure}

The next validation case is from the paper on the experimental test device Divertorlets \cite{saenz2022divertorlets}. This setup has been recreated using the FreeMHD simulations to further demonstrate the solver's main purpose in accurately modeling free-surface MHD flows.  The method for velocity measurements was using a pitot tube approach, where the difference in height in tubes was correlated to the velocity of the interior flow.  These main experimental measurements of the velocity between the vertical slats were compared to simulations, with the results shown in Figure \ref{fig:DivertorletsVerticalVelocity}. For a magnetic field strength of 0.2T, the externally applied current was varied from 400 to 900 A, with both experimental and simulated results showing an increase in velocity with increased applied current. 

\section{Discussion}
\label{section:Discussion} 

An extensive number of cases have been used to verify and validate FreeMHD. By beginning with verification cases for closed flow in ducts through comparing to analytical solutions, the model can be considered to be properly and accurately implemented. 
Next, the validation used a variety of experiments to determine the accuracy of the model compared to real-world results. The first case involved a pipe with 3D magnetic profiles, testing the ability of FreeMHD to calculate the current distribution that developed based on these conditions. Next, the dam breaking case tested the evolution of the free surface in a flow with a balance between surface tension and gravity. Lastly, the LMX-U and divertorlets cases were used to test free surface with magnetic fields. 

\begin{figure}
    \centering
    \begin{subfigure}[t]{0.99\columnwidth}
        \centering
        \includegraphics[width=\textwidth]{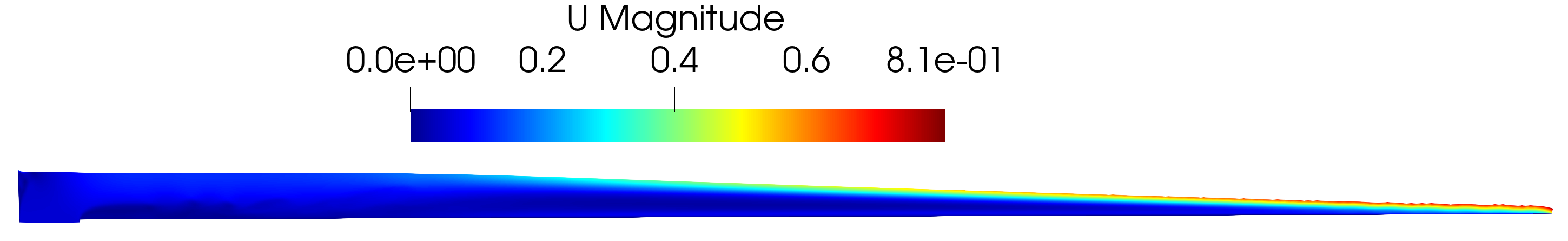}
        \caption{Streamwise Profile along Center of Channel, Velocity Magnitude (m/s)}
        \label{fig:lmx_0.3T_streamwise}
    \end{subfigure}
    \hfill
    
    \begin{subfigure}[t]{0.45\columnwidth}
        \centering
        \includegraphics[width=\textwidth]{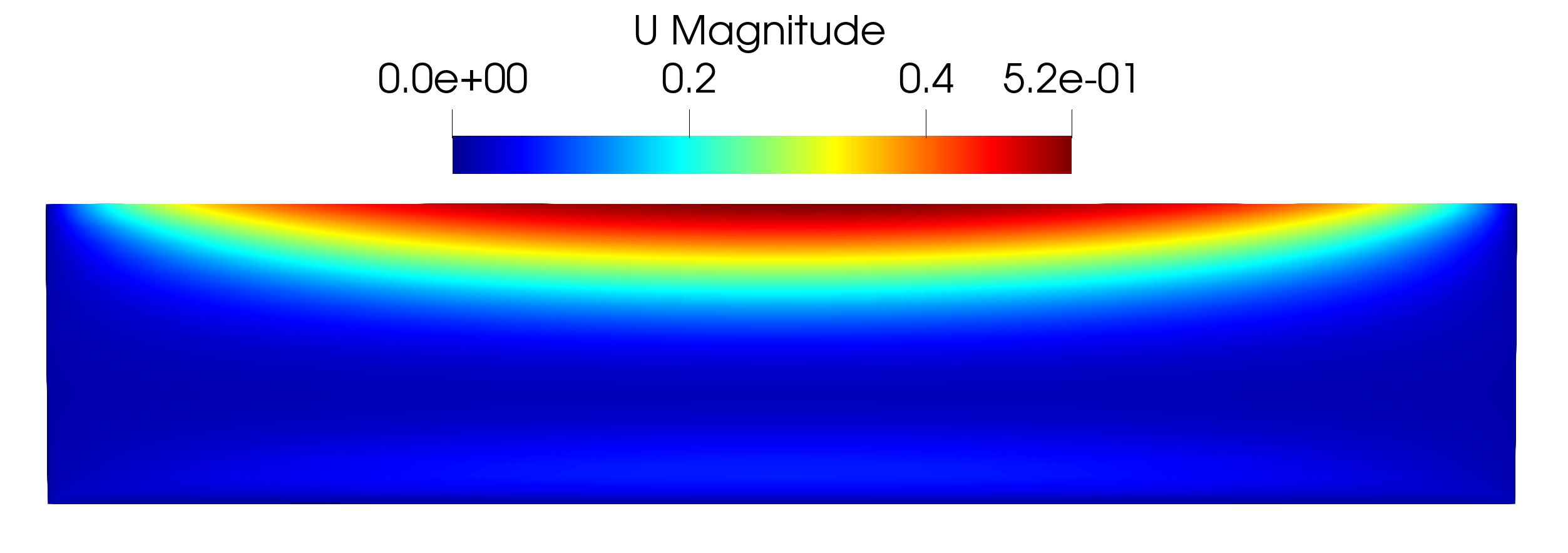}
        \caption{Cross Section Profile at Centerpoint of Magnet, Velocity Magnitude (m/s)}
        \label{fig:lmx_0.3T_XSlice0.37m_U}
    \end{subfigure}
    \hfill
    \begin{subfigure}[t]{0.45\columnwidth}
        \centering
        \includegraphics[width=\textwidth]{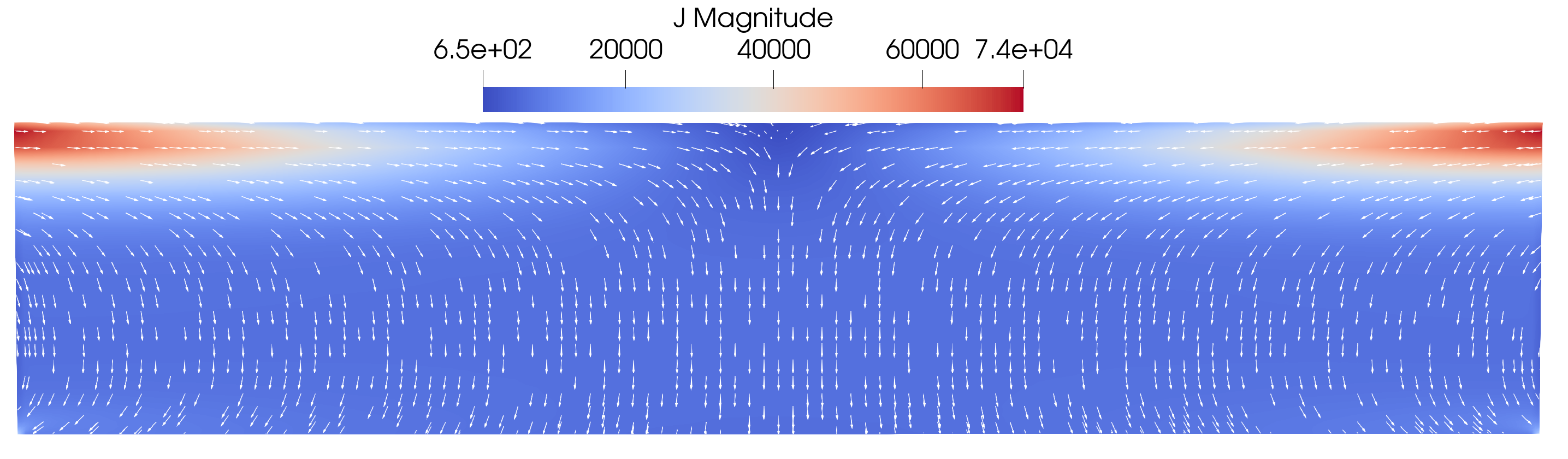}
        \caption{Cross Section Profile at Centerpoint of Magnet, Current Density Magnitude (A/m$^2$)}
        \label{fig:lmx_0.3T_XSlice0.37m_J}
    \end{subfigure}

    \caption{Profiles of LMX-U Results at 0.3T}
    \label{fig:LMX0.3TCase}
\end{figure}

The capabilities can be demonstrated with examples of velocity and current density profiles, as seen in Figure \ref{fig:LMX0.3TCase}. With the flow from left to right in the streamwise profile in Figure \ref{fig:lmx_0.3T_streamwise}, this shows the velocity over the height profiles previously shown in Figure \ref{fig:LMX_combinedBAndHeight}. This shows how the pileup occurs near the inlet due to MHD drag, and leads the flow to become thin and accelerated towards the outlet. The cross-section across the width of the duct at the midway of the magnet (x=0.37 m) in Figure \ref{fig:lmx_0.3T_XSlice0.37m_U}, further shows how the flow overall is opposed by the Lorentz force, but the flow along the top surface is accelerated. Lastly, at the same position, Figure \ref{fig:lmx_0.3T_XSlice0.37m_J} shows the current density induced by the flow across the magnetic field. Here the induced current density is primarily downwards, which then recirculates through the conductive liner.

Further applications involve extending current cases to more extreme conditions, such as higher flow speeds and magnetic field strengths and the application of external current for manipulation of flows.

\section{Conclusions}
\label{section:Conclusions}

FreeMHD has been validated through a series of cases, and these results confirm that FreeMHD is a reliable tool for designing LM systems under free surface conditions at the fusion reactor scale. Moreover, it is flexible, computationally efficient, and capable of solving fully 3D transient MHD flows.

Future developments for FreeMHD involve upgrades to further improve its versatility and applicability. While the currently used electric potential formulation is sufficient for the majority of liquid metal applications, some extreme conditions can necessitate magnetic field induction. Therefore, efforts are ongoing to implement a solver within FreeMHD involving an induction formulation.  
Additionally, the solver contains the equations and ability to calculate heat transfer effects, but this will be further validated in a following paper. Lastly, thermoelectric effects due to temperature gradients can be significant in a fusion environment, and will be implemented to enable further studies. 

\section{Funding Acknowledgment}
This research is supported by US DOE Grant No. DE-SC0024626. 

\section{Author Declarations}
The authors have no conflicts to disclose. 

\section{Data Availability}
The data that support the findings of this study are available on Princeton Data Commons.

\nocite{*}
\bibliography{aipsamp}

\begin{thebibliography}{42}%
\makeatletter
\providecommand \@ifxundefined [1]{%
 \@ifx{#1\undefined}
}%
\providecommand \@ifnum [1]{%
 \ifnum #1\expandafter \@firstoftwo
 \else \expandafter \@secondoftwo
 \fi
}%
\providecommand \@ifx [1]{%
 \ifx #1\expandafter \@firstoftwo
 \else \expandafter \@secondoftwo
 \fi
}%
\providecommand \natexlab [1]{#1}%
\providecommand \enquote  [1]{``#1''}%
\providecommand \bibnamefont  [1]{#1}%
\providecommand \bibfnamefont [1]{#1}%
\providecommand \citenamefont [1]{#1}%
\providecommand \href@noop [0]{\@secondoftwo}%
\providecommand \href [0]{\begingroup \@sanitize@url \@href}%
\providecommand \@href[1]{\@@startlink{#1}\@@href}%
\providecommand \@@href[1]{\endgroup#1\@@endlink}%
\providecommand \@sanitize@url [0]{\catcode `\\12\catcode `\$12\catcode `\&12\catcode `\#12\catcode `\^12\catcode `\_12\catcode `\%12\relax}%
\providecommand \@@startlink[1]{}%
\providecommand \@@endlink[0]{}%
\providecommand \url  [0]{\begingroup\@sanitize@url \@url }%
\providecommand \@url [1]{\endgroup\@href {#1}{\urlprefix }}%
\providecommand \urlprefix  [0]{URL }%
\providecommand \Eprint [0]{\href }%
\providecommand \doibase [0]{http://dx.doi.org/}%
\providecommand \selectlanguage [0]{\@gobble}%
\providecommand \bibinfo  [0]{\@secondoftwo}%
\providecommand \bibfield  [0]{\@secondoftwo}%
\providecommand \translation [1]{[#1]}%
\providecommand \BibitemOpen [0]{}%
\providecommand \bibitemStop [0]{}%
\providecommand \bibitemNoStop [0]{.\EOS\space}%
\providecommand \EOS [0]{\spacefactor3000\relax}%
\providecommand \BibitemShut  [1]{\csname bibitem#1\endcsname}%
\let\auto@bib@innerbib\@empty
\bibitem [{\citenamefont {Pitts}\ \emph {et~al.}(2019)\citenamefont {Pitts}, \citenamefont {Bonnin}, \citenamefont {Escourbiac}, \citenamefont {Frerichs}, \citenamefont {Gunn}, \citenamefont {Hirai}, \citenamefont {Kukushkin}, \citenamefont {Kaveeva}, \citenamefont {Miller}, \citenamefont {Moulton} \emph {et~al.}}]{pitts2019physics}%
  \BibitemOpen
  \bibfield  {author} {\bibinfo {author} {\bibfnamefont {R.~A.}\ \bibnamefont {Pitts}}, \bibinfo {author} {\bibfnamefont {X.}~\bibnamefont {Bonnin}}, \bibinfo {author} {\bibfnamefont {F.}~\bibnamefont {Escourbiac}}, \bibinfo {author} {\bibfnamefont {H.}~\bibnamefont {Frerichs}}, \bibinfo {author} {\bibfnamefont {J.}~\bibnamefont {Gunn}}, \bibinfo {author} {\bibfnamefont {T.}~\bibnamefont {Hirai}}, \bibinfo {author} {\bibfnamefont {A.}~\bibnamefont {Kukushkin}}, \bibinfo {author} {\bibfnamefont {E.}~\bibnamefont {Kaveeva}}, \bibinfo {author} {\bibfnamefont {M.}~\bibnamefont {Miller}}, \bibinfo {author} {\bibfnamefont {D.}~\bibnamefont {Moulton}},  \emph {et~al.},\ }\bibfield  {title} {\enquote {\bibinfo {title} {Physics basis for the first iter tungsten divertor},}\ }\href@noop {} {\bibfield  {journal} {\bibinfo  {journal} {Nuclear Materials and Energy}\ }\textbf {\bibinfo {volume} {20}},\ \bibinfo {pages} {100696} (\bibinfo {year} {2019})}\BibitemShut {NoStop}%
\bibitem [{\citenamefont {Kaita}(2019)}]{kaita2019fusion}%
  \BibitemOpen
  \bibfield  {author} {\bibinfo {author} {\bibfnamefont {R.}~\bibnamefont {Kaita}},\ }\bibfield  {title} {\enquote {\bibinfo {title} {Fusion applications for lithium: wall conditioning in magnetic confinement devices},}\ }\href@noop {} {\bibfield  {journal} {\bibinfo  {journal} {Plasma Physics and Controlled Fusion}\ }\textbf {\bibinfo {volume} {61}},\ \bibinfo {pages} {113001} (\bibinfo {year} {2019})}\BibitemShut {NoStop}%
\bibitem [{\citenamefont {Krasheninnikov}, \citenamefont {Zakharov},\ and\ \citenamefont {Pereverzev}(2003)}]{krasheninnikov2003lithium}%
  \BibitemOpen
  \bibfield  {author} {\bibinfo {author} {\bibfnamefont {S.}~\bibnamefont {Krasheninnikov}}, \bibinfo {author} {\bibfnamefont {L.}~\bibnamefont {Zakharov}}, \ and\ \bibinfo {author} {\bibfnamefont {G.}~\bibnamefont {Pereverzev}},\ }\bibfield  {title} {\enquote {\bibinfo {title} {On lithium walls and the performance of magnetic fusion devices},}\ }\href@noop {} {\bibfield  {journal} {\bibinfo  {journal} {Physics of Plasmas}\ }\textbf {\bibinfo {volume} {10}},\ \bibinfo {pages} {1678--1682} (\bibinfo {year} {2003})}\BibitemShut {NoStop}%
\bibitem [{\citenamefont {Boyle}\ \emph {et~al.}(2017)\citenamefont {Boyle}, \citenamefont {Majeski}, \citenamefont {Schmitt}, \citenamefont {Hansen}, \citenamefont {Kaita}, \citenamefont {Kubota}, \citenamefont {Lucia},\ and\ \citenamefont {Rognlien}}]{boyle2017observation}%
  \BibitemOpen
  \bibfield  {author} {\bibinfo {author} {\bibfnamefont {D.}~\bibnamefont {Boyle}}, \bibinfo {author} {\bibfnamefont {R.}~\bibnamefont {Majeski}}, \bibinfo {author} {\bibfnamefont {J.}~\bibnamefont {Schmitt}}, \bibinfo {author} {\bibfnamefont {C.}~\bibnamefont {Hansen}}, \bibinfo {author} {\bibfnamefont {R.}~\bibnamefont {Kaita}}, \bibinfo {author} {\bibfnamefont {S.}~\bibnamefont {Kubota}}, \bibinfo {author} {\bibfnamefont {M.}~\bibnamefont {Lucia}}, \ and\ \bibinfo {author} {\bibfnamefont {T.}~\bibnamefont {Rognlien}},\ }\bibfield  {title} {\enquote {\bibinfo {title} {Observation of flat electron temperature profiles in the lithium tokamak experiment},}\ }\href@noop {} {\bibfield  {journal} {\bibinfo  {journal} {Physical Review Letters}\ }\textbf {\bibinfo {volume} {119}},\ \bibinfo {pages} {015001} (\bibinfo {year} {2017})}\BibitemShut {NoStop}%
\bibitem [{\citenamefont {Sun}\ \emph {et~al.}(2023)\citenamefont {Sun}, \citenamefont {Al~Salami}, \citenamefont {Khodak}, \citenamefont {Saenz}, \citenamefont {Wynne}, \citenamefont {Maingi}, \citenamefont {Hanada}, \citenamefont {Hu},\ and\ \citenamefont {Kolemen}}]{sun2023magnetohydrodynamics}%
  \BibitemOpen
  \bibfield  {author} {\bibinfo {author} {\bibfnamefont {Z.}~\bibnamefont {Sun}}, \bibinfo {author} {\bibfnamefont {J.}~\bibnamefont {Al~Salami}}, \bibinfo {author} {\bibfnamefont {A.}~\bibnamefont {Khodak}}, \bibinfo {author} {\bibfnamefont {F.}~\bibnamefont {Saenz}}, \bibinfo {author} {\bibfnamefont {B.}~\bibnamefont {Wynne}}, \bibinfo {author} {\bibfnamefont {R.}~\bibnamefont {Maingi}}, \bibinfo {author} {\bibfnamefont {K.}~\bibnamefont {Hanada}}, \bibinfo {author} {\bibfnamefont {C.}~\bibnamefont {Hu}}, \ and\ \bibinfo {author} {\bibfnamefont {E.}~\bibnamefont {Kolemen}},\ }\bibfield  {title} {\enquote {\bibinfo {title} {Magnetohydrodynamics in free surface liquid metal flow relevant to plasma-facing components},}\ }\href@noop {} {\bibfield  {journal} {\bibinfo  {journal} {Nuclear Fusion}\ }\textbf {\bibinfo {volume} {63}},\ \bibinfo {pages} {076022} (\bibinfo {year} {2023})}\BibitemShut {NoStop}%
\bibitem [{\citenamefont {Nygren}\ \emph {et~al.}(2004)\citenamefont {Nygren}, \citenamefont {Rognlien}, \citenamefont {Rensink}, \citenamefont {Smolentsev}, \citenamefont {Youssef}, \citenamefont {Sawan}, \citenamefont {Merrill}, \citenamefont {Eberle}, \citenamefont {Fogarty}, \citenamefont {Nelson} \emph {et~al.}}]{nygren2004fusion}%
  \BibitemOpen
  \bibfield  {author} {\bibinfo {author} {\bibfnamefont {R.}~\bibnamefont {Nygren}}, \bibinfo {author} {\bibfnamefont {T.}~\bibnamefont {Rognlien}}, \bibinfo {author} {\bibfnamefont {M.}~\bibnamefont {Rensink}}, \bibinfo {author} {\bibfnamefont {S.}~\bibnamefont {Smolentsev}}, \bibinfo {author} {\bibfnamefont {M.}~\bibnamefont {Youssef}}, \bibinfo {author} {\bibfnamefont {M.}~\bibnamefont {Sawan}}, \bibinfo {author} {\bibfnamefont {B.}~\bibnamefont {Merrill}}, \bibinfo {author} {\bibfnamefont {C.}~\bibnamefont {Eberle}}, \bibinfo {author} {\bibfnamefont {P.}~\bibnamefont {Fogarty}}, \bibinfo {author} {\bibfnamefont {B.}~\bibnamefont {Nelson}},  \emph {et~al.},\ }\bibfield  {title} {\enquote {\bibinfo {title} {A fusion reactor design with a liquid first wall and divertor},}\ }\href@noop {} {\bibfield  {journal} {\bibinfo  {journal} {Fusion Engineering and Design}\ }\textbf {\bibinfo {volume} {72}},\ \bibinfo {pages} {181--221} (\bibinfo {year} {2004})}\BibitemShut {NoStop}%
\bibitem [{\citenamefont {Gao}, \citenamefont {Morley},\ and\ \citenamefont {Dhir}(2002)}]{gao2002numerical}%
  \BibitemOpen
  \bibfield  {author} {\bibinfo {author} {\bibfnamefont {D.}~\bibnamefont {Gao}}, \bibinfo {author} {\bibfnamefont {N.}~\bibnamefont {Morley}}, \ and\ \bibinfo {author} {\bibfnamefont {V.}~\bibnamefont {Dhir}},\ }\bibfield  {title} {\enquote {\bibinfo {title} {Numerical study of liquid metal film flows in a varying spanwise magnetic field},}\ }\href@noop {} {\bibfield  {journal} {\bibinfo  {journal} {Fusion engineering and design}\ }\textbf {\bibinfo {volume} {63}},\ \bibinfo {pages} {369--374} (\bibinfo {year} {2002})}\BibitemShut {NoStop}%
\bibitem [{\citenamefont {Huang}, \citenamefont {Ying},\ and\ \citenamefont {Abdou}(2002)}]{huang20023d}%
  \BibitemOpen
  \bibfield  {author} {\bibinfo {author} {\bibfnamefont {H.}~\bibnamefont {Huang}}, \bibinfo {author} {\bibfnamefont {A.}~\bibnamefont {Ying}}, \ and\ \bibinfo {author} {\bibfnamefont {M.}~\bibnamefont {Abdou}},\ }\bibfield  {title} {\enquote {\bibinfo {title} {3d mhd free surface fluid flow simulation based on magnetic-field induction equations},}\ }\href@noop {} {\bibfield  {journal} {\bibinfo  {journal} {Fusion Engineering and Design}\ }\textbf {\bibinfo {volume} {63}},\ \bibinfo {pages} {361--368} (\bibinfo {year} {2002})}\BibitemShut {NoStop}%
\bibitem [{\citenamefont {Morley}\ \emph {et~al.}(2004)\citenamefont {Morley}, \citenamefont {Smolentsev}, \citenamefont {Munipalli}, \citenamefont {Ni}, \citenamefont {Gao},\ and\ \citenamefont {Abdou}}]{morley2004progress}%
  \BibitemOpen
  \bibfield  {author} {\bibinfo {author} {\bibfnamefont {N.}~\bibnamefont {Morley}}, \bibinfo {author} {\bibfnamefont {S.}~\bibnamefont {Smolentsev}}, \bibinfo {author} {\bibfnamefont {R.}~\bibnamefont {Munipalli}}, \bibinfo {author} {\bibfnamefont {M.-J.}\ \bibnamefont {Ni}}, \bibinfo {author} {\bibfnamefont {D.}~\bibnamefont {Gao}}, \ and\ \bibinfo {author} {\bibfnamefont {M.}~\bibnamefont {Abdou}},\ }\bibfield  {title} {\enquote {\bibinfo {title} {Progress on the modeling of liquid metal, free surface, mhd flows for fusion liquid walls},}\ }\href@noop {} {\bibfield  {journal} {\bibinfo  {journal} {Fusion Engineering and Design}\ }\textbf {\bibinfo {volume} {72}},\ \bibinfo {pages} {3--34} (\bibinfo {year} {2004})}\BibitemShut {NoStop}%
\bibitem [{\citenamefont {Siriano}\ \emph {et~al.}(2024)\citenamefont {Siriano}, \citenamefont {Melchiorri}, \citenamefont {Pignatiello},\ and\ \citenamefont {Tassone}}]{siriano2024multi}%
  \BibitemOpen
  \bibfield  {author} {\bibinfo {author} {\bibfnamefont {S.}~\bibnamefont {Siriano}}, \bibinfo {author} {\bibfnamefont {L.}~\bibnamefont {Melchiorri}}, \bibinfo {author} {\bibfnamefont {S.}~\bibnamefont {Pignatiello}}, \ and\ \bibinfo {author} {\bibfnamefont {A.}~\bibnamefont {Tassone}},\ }\bibfield  {title} {\enquote {\bibinfo {title} {A multi-region and a multiphase mhd openfoam solver for fusion reactor analysis},}\ }\href@noop {} {\bibfield  {journal} {\bibinfo  {journal} {Fusion Engineering and Design}\ }\textbf {\bibinfo {volume} {200}},\ \bibinfo {pages} {114216} (\bibinfo {year} {2024})}\BibitemShut {NoStop}%
\bibitem [{\citenamefont {Salami}(2022)}]{AlSalami2022numerical}%
  \BibitemOpen
  \bibfield  {author} {\bibinfo {author} {\bibfnamefont {J.~S. S.~A.}\ \bibnamefont {Salami}},\ }\bibfield  {title} {\enquote {\bibinfo {title} {Numerical simulation of mhd free surface liquid metal flows for nuclear fusion applications},}\ }\href@noop {} {\bibfield  {journal} {\bibinfo  {journal} {(Kyuhsu University)}\ } (\bibinfo {year} {2022})}\BibitemShut {NoStop}%
\bibitem [{Note1()}]{Note1}%
  \BibitemOpen
  \bibinfo {note} {FreeMHD \protect \url {https://github.com/PlasmaControl/FreeMHD}.}\BibitemShut {Stop}%
\bibitem [{\citenamefont {Brackbill}, \citenamefont {Kothe},\ and\ \citenamefont {Zemach}(1992)}]{brackbill1992continuum}%
  \BibitemOpen
  \bibfield  {author} {\bibinfo {author} {\bibfnamefont {J.~U.}\ \bibnamefont {Brackbill}}, \bibinfo {author} {\bibfnamefont {D.~B.}\ \bibnamefont {Kothe}}, \ and\ \bibinfo {author} {\bibfnamefont {C.}~\bibnamefont {Zemach}},\ }\bibfield  {title} {\enquote {\bibinfo {title} {A continuum method for modeling surface tension},}\ }\href@noop {} {\bibfield  {journal} {\bibinfo  {journal} {Journal of computational physics}\ }\textbf {\bibinfo {volume} {100}},\ \bibinfo {pages} {335--354} (\bibinfo {year} {1992})}\BibitemShut {NoStop}%
\bibitem [{\citenamefont {Moreau}(1990)}]{moreau1990magnetohydrodynamics}%
  \BibitemOpen
  \bibfield  {author} {\bibinfo {author} {\bibfnamefont {R.}~\bibnamefont {Moreau}},\ }\href@noop {} {\emph {\bibinfo {title} {Magnetohydrodynamics}}},\ Vol.~\bibinfo {volume} {3}\ (\bibinfo  {publisher} {Kluwer Academic Publisher},\ \bibinfo {year} {1990})\BibitemShut {NoStop}%
\bibitem [{\citenamefont {M{\"u}ller}\ and\ \citenamefont {B{\"u}hler}(2001)}]{muller2001magnetofluiddynamics}%
  \BibitemOpen
  \bibfield  {author} {\bibinfo {author} {\bibfnamefont {U.}~\bibnamefont {M{\"u}ller}}\ and\ \bibinfo {author} {\bibfnamefont {L.}~\bibnamefont {B{\"u}hler}},\ }\href@noop {} {\emph {\bibinfo {title} {Magnetofluiddynamics in channels and containers}}}\ (\bibinfo  {publisher} {Springer Science \& Business Media},\ \bibinfo {year} {2001})\BibitemShut {NoStop}%
\bibitem [{\citenamefont {Ni}\ \emph {et~al.}(2007)\citenamefont {Ni}, \citenamefont {Munipalli}, \citenamefont {Huang}, \citenamefont {Morley},\ and\ \citenamefont {Abdou}}]{ni2007current}%
  \BibitemOpen
  \bibfield  {author} {\bibinfo {author} {\bibfnamefont {M.-J.}\ \bibnamefont {Ni}}, \bibinfo {author} {\bibfnamefont {R.}~\bibnamefont {Munipalli}}, \bibinfo {author} {\bibfnamefont {P.}~\bibnamefont {Huang}}, \bibinfo {author} {\bibfnamefont {N.~B.}\ \bibnamefont {Morley}}, \ and\ \bibinfo {author} {\bibfnamefont {M.~A.}\ \bibnamefont {Abdou}},\ }\bibfield  {title} {\enquote {\bibinfo {title} {A current density conservative scheme for incompressible mhd flows at a low magnetic reynolds number. part ii: On an arbitrary collocated mesh},}\ }\href@noop {} {\bibfield  {journal} {\bibinfo  {journal} {Journal of Computational Physics}\ }\textbf {\bibinfo {volume} {227}},\ \bibinfo {pages} {205--228} (\bibinfo {year} {2007})}\BibitemShut {NoStop}%
\bibitem [{\citenamefont {Morley}\ and\ \citenamefont {Burris}(2003)}]{morley2003mtor}%
  \BibitemOpen
  \bibfield  {author} {\bibinfo {author} {\bibfnamefont {N.~B.}\ \bibnamefont {Morley}}\ and\ \bibinfo {author} {\bibfnamefont {J.}~\bibnamefont {Burris}},\ }\bibfield  {title} {\enquote {\bibinfo {title} {The mtor lm-mhd flow facility, and preliminary experimental investigation of thin layer, liquid metal flow in a 1/r toroidal magnetic field},}\ }\href@noop {} {\bibfield  {journal} {\bibinfo  {journal} {Fusion science and technology}\ }\textbf {\bibinfo {volume} {44}},\ \bibinfo {pages} {74--78} (\bibinfo {year} {2003})}\BibitemShut {NoStop}%
\bibitem [{\citenamefont {O'Donnell}, \citenamefont {Papanikolaou},\ and\ \citenamefont {Reed}(1989)}]{ODonnell1989thermophysical}%
  \BibitemOpen
  \bibfield  {author} {\bibinfo {author} {\bibfnamefont {W.~J.}\ \bibnamefont {O'Donnell}}, \bibinfo {author} {\bibfnamefont {P.~G.}\ \bibnamefont {Papanikolaou}}, \ and\ \bibinfo {author} {\bibfnamefont {C.~B.}\ \bibnamefont {Reed}},\ }\href@noop {} {\enquote {\bibinfo {title} {The thermophysical and transport properties of eutectic nak near room temperature},}\ }\bibinfo {type} {Tech. Rep.}\ (\bibinfo  {institution} {Argonne National Lab.},\ \bibinfo {year} {1989})\BibitemShut {NoStop}%
\bibitem [{\citenamefont {Moresco}\ and\ \citenamefont {Alboussiere}(2004)}]{moresco2004experimental}%
  \BibitemOpen
  \bibfield  {author} {\bibinfo {author} {\bibfnamefont {P.}~\bibnamefont {Moresco}}\ and\ \bibinfo {author} {\bibfnamefont {T.}~\bibnamefont {Alboussiere}},\ }\bibfield  {title} {\enquote {\bibinfo {title} {Experimental study of the instability of the hartmann layer},}\ }\href@noop {} {\bibfield  {journal} {\bibinfo  {journal} {Journal of Fluid Mechanics}\ }\textbf {\bibinfo {volume} {504}},\ \bibinfo {pages} {167--181} (\bibinfo {year} {2004})}\BibitemShut {NoStop}%
\bibitem [{\citenamefont {Farhadi}\ \emph {et~al.}(2018)\citenamefont {Farhadi}, \citenamefont {Mayrhofer}, \citenamefont {Tritthart}, \citenamefont {Glas},\ and\ \citenamefont {Habersack}}]{farhadi2018accuracy}%
  \BibitemOpen
  \bibfield  {author} {\bibinfo {author} {\bibfnamefont {A.}~\bibnamefont {Farhadi}}, \bibinfo {author} {\bibfnamefont {A.}~\bibnamefont {Mayrhofer}}, \bibinfo {author} {\bibfnamefont {M.}~\bibnamefont {Tritthart}}, \bibinfo {author} {\bibfnamefont {M.}~\bibnamefont {Glas}}, \ and\ \bibinfo {author} {\bibfnamefont {H.}~\bibnamefont {Habersack}},\ }\bibfield  {title} {\enquote {\bibinfo {title} {Accuracy and comparison of standard k-$\epsilon$ with two variants of k-$\omega$ turbulence models in fluvial applications},}\ }\href@noop {} {\bibfield  {journal} {\bibinfo  {journal} {Engineering Applications of Computational Fluid Mechanics}\ }\textbf {\bibinfo {volume} {12}},\ \bibinfo {pages} {216--235} (\bibinfo {year} {2018})}\BibitemShut {NoStop}%
\bibitem [{\citenamefont {Shaheed}, \citenamefont {Mohammadian},\ and\ \citenamefont {Kheirkhah~Gildeh}(2019)}]{shaheed2019comparison}%
  \BibitemOpen
  \bibfield  {author} {\bibinfo {author} {\bibfnamefont {R.}~\bibnamefont {Shaheed}}, \bibinfo {author} {\bibfnamefont {A.}~\bibnamefont {Mohammadian}}, \ and\ \bibinfo {author} {\bibfnamefont {H.}~\bibnamefont {Kheirkhah~Gildeh}},\ }\bibfield  {title} {\enquote {\bibinfo {title} {A comparison of standard k--$\varepsilon$ and realizable k--$\varepsilon$ turbulence models in curved and confluent channels},}\ }\href@noop {} {\bibfield  {journal} {\bibinfo  {journal} {Environmental Fluid Mechanics}\ }\textbf {\bibinfo {volume} {19}},\ \bibinfo {pages} {543--568} (\bibinfo {year} {2019})}\BibitemShut {NoStop}%
\bibitem [{\citenamefont {Shercliff}(1953)}]{shercliff1953steady}%
  \BibitemOpen
  \bibfield  {author} {\bibinfo {author} {\bibfnamefont {J.}~\bibnamefont {Shercliff}},\ }\bibfield  {title} {\enquote {\bibinfo {title} {Steady motion of conducting fluids in pipes under transverse magnetic fields},}\ }in\ \href@noop {} {\emph {\bibinfo {booktitle} {Mathematical Proceedings of the Cambridge Philosophical Society}}},\ Vol.~\bibinfo {volume} {49}\ (\bibinfo {organization} {Cambridge University Press},\ \bibinfo {year} {1953})\ pp.\ \bibinfo {pages} {136--144}\BibitemShut {NoStop}%
\bibitem [{\citenamefont {Hunt}(1965)}]{hunt1965magnetohydrodynamic}%
  \BibitemOpen
  \bibfield  {author} {\bibinfo {author} {\bibfnamefont {J.}~\bibnamefont {Hunt}},\ }\bibfield  {title} {\enquote {\bibinfo {title} {Magnetohydrodynamic flow in rectangular ducts},}\ }\href@noop {} {\bibfield  {journal} {\bibinfo  {journal} {Journal of fluid mechanics}\ }\textbf {\bibinfo {volume} {21}},\ \bibinfo {pages} {577--590} (\bibinfo {year} {1965})}\BibitemShut {NoStop}%
\bibitem [{\citenamefont {B{\"u}hler}, \citenamefont {Brinkmann},\ and\ \citenamefont {Mistrangelo}(2020)}]{buhler2020experimental}%
  \BibitemOpen
  \bibfield  {author} {\bibinfo {author} {\bibfnamefont {L.}~\bibnamefont {B{\"u}hler}}, \bibinfo {author} {\bibfnamefont {H.-J.}\ \bibnamefont {Brinkmann}}, \ and\ \bibinfo {author} {\bibfnamefont {C.}~\bibnamefont {Mistrangelo}},\ }\bibfield  {title} {\enquote {\bibinfo {title} {Experimental investigation of liquid metal pipe flow in a strong non-uniform magnetic field.}}\ }\href@noop {} {\bibfield  {journal} {\bibinfo  {journal} {Magnetohydrodynamics (0024-998X)}\ }\textbf {\bibinfo {volume} {56}} (\bibinfo {year} {2020})}\BibitemShut {NoStop}%
\bibitem [{\citenamefont {Ozmen-Cagatay}\ and\ \citenamefont {Kocaman}(2010)}]{ozmen2010dam}%
  \BibitemOpen
  \bibfield  {author} {\bibinfo {author} {\bibfnamefont {H.}~\bibnamefont {Ozmen-Cagatay}}\ and\ \bibinfo {author} {\bibfnamefont {S.}~\bibnamefont {Kocaman}},\ }\bibfield  {title} {\enquote {\bibinfo {title} {Dam-break flows during initial stage using swe and rans approaches},}\ }\href@noop {} {\bibfield  {journal} {\bibinfo  {journal} {Journal of Hydraulic Research}\ }\textbf {\bibinfo {volume} {48}},\ \bibinfo {pages} {603--611} (\bibinfo {year} {2010})}\BibitemShut {NoStop}%
\bibitem [{\citenamefont {Saenz}\ \emph {et~al.}(2022)\citenamefont {Saenz}, \citenamefont {Sun}, \citenamefont {Fisher}, \citenamefont {Wynne},\ and\ \citenamefont {Kolemen}}]{saenz2022divertorlets}%
  \BibitemOpen
  \bibfield  {author} {\bibinfo {author} {\bibfnamefont {F.}~\bibnamefont {Saenz}}, \bibinfo {author} {\bibfnamefont {Z.}~\bibnamefont {Sun}}, \bibinfo {author} {\bibfnamefont {A.~E.}\ \bibnamefont {Fisher}}, \bibinfo {author} {\bibfnamefont {B.}~\bibnamefont {Wynne}}, \ and\ \bibinfo {author} {\bibfnamefont {E.}~\bibnamefont {Kolemen}},\ }\bibfield  {title} {\enquote {\bibinfo {title} {Divertorlets concept for low-recycling fusion reactor divertor: experimental, analytical and numerical verification},}\ }\href@noop {} {\bibfield  {journal} {\bibinfo  {journal} {Nuclear Fusion}\ }\textbf {\bibinfo {volume} {62}},\ \bibinfo {pages} {086008} (\bibinfo {year} {2022})}\BibitemShut {NoStop}%
\bibitem [{\citenamefont {Shah}, \citenamefont {London},\ and\ \citenamefont {White}(1980)}]{shah1980laminar}%
  \BibitemOpen
  \bibfield  {author} {\bibinfo {author} {\bibfnamefont {R.~K.}\ \bibnamefont {Shah}}, \bibinfo {author} {\bibfnamefont {A.}~\bibnamefont {London}}, \ and\ \bibinfo {author} {\bibfnamefont {F.~M.}\ \bibnamefont {White}},\ }\bibfield  {title} {\enquote {\bibinfo {title} {Laminar flow forced convection in ducts},}\ }\href@noop {} {\  (\bibinfo {year} {1980})}\BibitemShut {NoStop}%
\bibitem [{\citenamefont {Munipalli}\ \emph {et~al.}(2003)\citenamefont {Munipalli}, \citenamefont {Shankar}, \citenamefont {Chandler}, \citenamefont {Rowell}, \citenamefont {Ni}, \citenamefont {Smolentsev}, \citenamefont {Morley}, \citenamefont {Abdou},\ and\ \citenamefont {Hadid}}]{ramakanth2003development}%
  \BibitemOpen
  \bibfield  {author} {\bibinfo {author} {\bibfnamefont {R.}~\bibnamefont {Munipalli}}, \bibinfo {author} {\bibfnamefont {V.}~\bibnamefont {Shankar}}, \bibinfo {author} {\bibfnamefont {C.}~\bibnamefont {Chandler}}, \bibinfo {author} {\bibfnamefont {C.}~\bibnamefont {Rowell}}, \bibinfo {author} {\bibfnamefont {M.}~\bibnamefont {Ni}}, \bibinfo {author} {\bibfnamefont {S.}~\bibnamefont {Smolentsev}}, \bibinfo {author} {\bibfnamefont {N.}~\bibnamefont {Morley}}, \bibinfo {author} {\bibfnamefont {M.}~\bibnamefont {Abdou}}, \ and\ \bibinfo {author} {\bibfnamefont {A.}~\bibnamefont {Hadid}},\ }\href@noop {} {\enquote {\bibinfo {title} {Development of a 3-d incompressible free surface mhd computational environment for arbitrary geometries: Himag},}\ }\bibinfo {type} {Tech. Rep.}\ (\bibinfo  {institution} {HyPerComp Inc.},\ \bibinfo {year} {2003})\BibitemShut {NoStop}%
\bibitem [{\citenamefont {Moukalled}\ \emph {et~al.}(2016)\citenamefont {Moukalled}, \citenamefont {Mangani}, \citenamefont {Darwish}, \citenamefont {Moukalled}, \citenamefont {Mangani},\ and\ \citenamefont {Darwish}}]{moukalled2016finite}%
  \BibitemOpen
  \bibfield  {author} {\bibinfo {author} {\bibfnamefont {F.}~\bibnamefont {Moukalled}}, \bibinfo {author} {\bibfnamefont {L.}~\bibnamefont {Mangani}}, \bibinfo {author} {\bibfnamefont {M.}~\bibnamefont {Darwish}}, \bibinfo {author} {\bibfnamefont {F.}~\bibnamefont {Moukalled}}, \bibinfo {author} {\bibfnamefont {L.}~\bibnamefont {Mangani}}, \ and\ \bibinfo {author} {\bibfnamefont {M.}~\bibnamefont {Darwish}},\ }\href@noop {} {\emph {\bibinfo {title} {The finite volume method}}}\ (\bibinfo  {publisher} {Springer},\ \bibinfo {year} {2016})\BibitemShut {NoStop}%
\bibitem [{\citenamefont {Darwish}\ and\ \citenamefont {Moukalled}(2003)}]{darwish2003tvd}%
  \BibitemOpen
  \bibfield  {author} {\bibinfo {author} {\bibfnamefont {M.}~\bibnamefont {Darwish}}\ and\ \bibinfo {author} {\bibfnamefont {F.}~\bibnamefont {Moukalled}},\ }\bibfield  {title} {\enquote {\bibinfo {title} {Tvd schemes for unstructured grids},}\ }\href@noop {} {\bibfield  {journal} {\bibinfo  {journal} {International Journal of heat and mass transfer}\ }\textbf {\bibinfo {volume} {46}},\ \bibinfo {pages} {599--611} (\bibinfo {year} {2003})}\BibitemShut {NoStop}%
\bibitem [{\citenamefont {Picologlou}\ and\ \citenamefont {Reed}(1989)}]{picologlou1989experimental}%
  \BibitemOpen
  \bibfield  {author} {\bibinfo {author} {\bibfnamefont {B.}~\bibnamefont {Picologlou}}\ and\ \bibinfo {author} {\bibfnamefont {C.}~\bibnamefont {Reed}},\ }\bibfield  {title} {\enquote {\bibinfo {title} {Experimental investigation of 3-d mhd flows at high hartmann number and interaction parameter},}\ }in\ \href@noop {} {\emph {\bibinfo {booktitle} {Liquid Metal Magnetohydrodynamics}}}\ (\bibinfo  {publisher} {Springer},\ \bibinfo {year} {1989})\ pp.\ \bibinfo {pages} {71--77}\BibitemShut {NoStop}%
\bibitem [{\citenamefont {Reed}\ \emph {et~al.}(1987)\citenamefont {Reed}, \citenamefont {Picologlou}, \citenamefont {Hua},\ and\ \citenamefont {Walker}}]{reed1987alex}%
  \BibitemOpen
  \bibfield  {author} {\bibinfo {author} {\bibfnamefont {C.}~\bibnamefont {Reed}}, \bibinfo {author} {\bibfnamefont {B.}~\bibnamefont {Picologlou}}, \bibinfo {author} {\bibfnamefont {T.}~\bibnamefont {Hua}}, \ and\ \bibinfo {author} {\bibfnamefont {J.}~\bibnamefont {Walker}},\ }\href@noop {} {\enquote {\bibinfo {title} {Alex results: A comparison of measurements from a round and a rectangular duct with 3-d code predictions},}\ }\bibinfo {type} {Tech. Rep.}\ (\bibinfo  {institution} {Argonne National Lab.},\ \bibinfo {year} {1987})\BibitemShut {NoStop}%
\bibitem [{\citenamefont {Smolentsev}\ \emph {et~al.}(2015)\citenamefont {Smolentsev}, \citenamefont {Badia}, \citenamefont {Bhattacharyay}, \citenamefont {B{\"u}hler}, \citenamefont {Chen}, \citenamefont {Huang}, \citenamefont {Jin}, \citenamefont {Krasnov}, \citenamefont {Lee}, \citenamefont {De~Les~Valls} \emph {et~al.}}]{smolentsev2015approach}%
  \BibitemOpen
  \bibfield  {author} {\bibinfo {author} {\bibfnamefont {S.}~\bibnamefont {Smolentsev}}, \bibinfo {author} {\bibfnamefont {S.}~\bibnamefont {Badia}}, \bibinfo {author} {\bibfnamefont {R.}~\bibnamefont {Bhattacharyay}}, \bibinfo {author} {\bibfnamefont {L.}~\bibnamefont {B{\"u}hler}}, \bibinfo {author} {\bibfnamefont {L.}~\bibnamefont {Chen}}, \bibinfo {author} {\bibfnamefont {Q.}~\bibnamefont {Huang}}, \bibinfo {author} {\bibfnamefont {H.-G.}\ \bibnamefont {Jin}}, \bibinfo {author} {\bibfnamefont {D.}~\bibnamefont {Krasnov}}, \bibinfo {author} {\bibfnamefont {D.-W.}\ \bibnamefont {Lee}}, \bibinfo {author} {\bibfnamefont {E.~M.}\ \bibnamefont {De~Les~Valls}},  \emph {et~al.},\ }\bibfield  {title} {\enquote {\bibinfo {title} {An approach to verification and validation of mhd codes for fusion applications},}\ }\href@noop {} {\bibfield  {journal} {\bibinfo  {journal} {Fusion Engineering and Design}\ }\textbf {\bibinfo {volume} {100}},\ \bibinfo {pages} {65--72} (\bibinfo {year} {2015})}\BibitemShut {NoStop}%
\bibitem [{\citenamefont {Davison}(1968)}]{davison1968compilation}%
  \BibitemOpen
  \bibfield  {author} {\bibinfo {author} {\bibfnamefont {H.~W.}\ \bibnamefont {Davison}},\ }\href@noop {} {\emph {\bibinfo {title} {Compilation of thermophysical properties of liquid lithium}}},\ Vol.\ \bibinfo {volume} {4650}\ (\bibinfo  {publisher} {National Aeronautics and Space Administration},\ \bibinfo {year} {1968})\BibitemShut {NoStop}%
\bibitem [{\citenamefont {Ying}\ \emph {et~al.}(2004)\citenamefont {Ying}, \citenamefont {Abdou}, \citenamefont {Morley}, \citenamefont {Sketchley}, \citenamefont {Woolley}, \citenamefont {Burris}, \citenamefont {Kaita}, \citenamefont {Fogarty}, \citenamefont {Huang}, \citenamefont {Lao} \emph {et~al.}}]{ying2004exploratory}%
  \BibitemOpen
  \bibfield  {author} {\bibinfo {author} {\bibfnamefont {A.}~\bibnamefont {Ying}}, \bibinfo {author} {\bibfnamefont {M.}~\bibnamefont {Abdou}}, \bibinfo {author} {\bibfnamefont {N.}~\bibnamefont {Morley}}, \bibinfo {author} {\bibfnamefont {T.}~\bibnamefont {Sketchley}}, \bibinfo {author} {\bibfnamefont {R.}~\bibnamefont {Woolley}}, \bibinfo {author} {\bibfnamefont {J.}~\bibnamefont {Burris}}, \bibinfo {author} {\bibfnamefont {R.}~\bibnamefont {Kaita}}, \bibinfo {author} {\bibfnamefont {P.}~\bibnamefont {Fogarty}}, \bibinfo {author} {\bibfnamefont {H.}~\bibnamefont {Huang}}, \bibinfo {author} {\bibfnamefont {X.}~\bibnamefont {Lao}},  \emph {et~al.},\ }\bibfield  {title} {\enquote {\bibinfo {title} {Exploratory studies of flowing liquid metal divertor options for fusion-relevant magnetic fields in the mtor facility},}\ }\href@noop {} {\bibfield  {journal} {\bibinfo  {journal} {Fusion engineering and design}\ }\textbf {\bibinfo {volume} {72}},\ \bibinfo {pages} {35--62} (\bibinfo {year} {2004})}\BibitemShut
  {NoStop}%
\bibitem [{\citenamefont {Saenz}\ \emph {et~al.}(2023)\citenamefont {Saenz}, \citenamefont {Fisher}, \citenamefont {Al-Salami}, \citenamefont {Wynne}, \citenamefont {Sun}, \citenamefont {Tanaka}, \citenamefont {Kunugi}, \citenamefont {Yagi}, \citenamefont {Kusumi}, \citenamefont {Wu} \emph {et~al.}}]{saenz2023experimental}%
  \BibitemOpen
  \bibfield  {author} {\bibinfo {author} {\bibfnamefont {F.}~\bibnamefont {Saenz}}, \bibinfo {author} {\bibfnamefont {A.}~\bibnamefont {Fisher}}, \bibinfo {author} {\bibfnamefont {J.}~\bibnamefont {Al-Salami}}, \bibinfo {author} {\bibfnamefont {B.}~\bibnamefont {Wynne}}, \bibinfo {author} {\bibfnamefont {Z.}~\bibnamefont {Sun}}, \bibinfo {author} {\bibfnamefont {T.}~\bibnamefont {Tanaka}}, \bibinfo {author} {\bibfnamefont {T.}~\bibnamefont {Kunugi}}, \bibinfo {author} {\bibfnamefont {J.}~\bibnamefont {Yagi}}, \bibinfo {author} {\bibfnamefont {K.}~\bibnamefont {Kusumi}}, \bibinfo {author} {\bibfnamefont {Y.}~\bibnamefont {Wu}},  \emph {et~al.},\ }\bibfield  {title} {\enquote {\bibinfo {title} {Experimental, numerical and analytical evaluation of-thrust for fast-liquid-metal-flow divertor systems of nuclear fusion devices},}\ }\href@noop {} {\bibfield  {journal} {\bibinfo  {journal} {Nuclear Fusion}\ }\textbf {\bibinfo {volume} {63}},\ \bibinfo {pages} {096015} (\bibinfo {year} {2023})}\BibitemShut {NoStop}%
\bibitem [{\citenamefont {Davidson}(2001)}]{davidson2001introduction}%
  \BibitemOpen
  \bibfield  {author} {\bibinfo {author} {\bibfnamefont {P.~A.}\ \bibnamefont {Davidson}},\ }\href@noop {} {\emph {\bibinfo {title} {An Introduction to Magnetohydrodynamics}}},\ Cambridge Texts in Applied Mathematics\ (\bibinfo  {publisher} {Cambridge University Press},\ \bibinfo {year} {2001})\BibitemShut {NoStop}%
\bibitem [{\citenamefont {Loarte}\ \emph {et~al.}(2007)\citenamefont {Loarte}, \citenamefont {Lipschultz}, \citenamefont {Kukushkin}, \citenamefont {Matthews}, \citenamefont {Stangeby}, \citenamefont {Asakura}, \citenamefont {Counsell}, \citenamefont {Federici}, \citenamefont {Kallenbach}, \citenamefont {Krieger} \emph {et~al.}}]{loarte2007power}%
  \BibitemOpen
  \bibfield  {author} {\bibinfo {author} {\bibfnamefont {A.}~\bibnamefont {Loarte}}, \bibinfo {author} {\bibfnamefont {B.}~\bibnamefont {Lipschultz}}, \bibinfo {author} {\bibfnamefont {A.}~\bibnamefont {Kukushkin}}, \bibinfo {author} {\bibfnamefont {G.}~\bibnamefont {Matthews}}, \bibinfo {author} {\bibfnamefont {P.}~\bibnamefont {Stangeby}}, \bibinfo {author} {\bibfnamefont {N.}~\bibnamefont {Asakura}}, \bibinfo {author} {\bibfnamefont {G.}~\bibnamefont {Counsell}}, \bibinfo {author} {\bibfnamefont {G.}~\bibnamefont {Federici}}, \bibinfo {author} {\bibfnamefont {A.}~\bibnamefont {Kallenbach}}, \bibinfo {author} {\bibfnamefont {K.}~\bibnamefont {Krieger}},  \emph {et~al.},\ }\bibfield  {title} {\enquote {\bibinfo {title} {Power and particle control},}\ }\href@noop {} {\bibfield  {journal} {\bibinfo  {journal} {Nuclear Fusion}\ }\textbf {\bibinfo {volume} {47}},\ \bibinfo {pages} {S203} (\bibinfo {year} {2007})}\BibitemShut {NoStop}%
\bibitem [{\citenamefont {Greenwald}\ \emph {et~al.}(2007)\citenamefont {Greenwald}, \citenamefont {Callis}, \citenamefont {Gates}, \citenamefont {Dorland}, \citenamefont {Harris}, \citenamefont {Linford}, \citenamefont {Mauel}, \citenamefont {McCarthy}, \citenamefont {Meade}, \citenamefont {Najmabadi} \emph {et~al.}}]{greenwald2007priorities}%
  \BibitemOpen
  \bibfield  {author} {\bibinfo {author} {\bibfnamefont {M.}~\bibnamefont {Greenwald}}, \bibinfo {author} {\bibfnamefont {R.}~\bibnamefont {Callis}}, \bibinfo {author} {\bibfnamefont {D.}~\bibnamefont {Gates}}, \bibinfo {author} {\bibfnamefont {W.}~\bibnamefont {Dorland}}, \bibinfo {author} {\bibfnamefont {J.}~\bibnamefont {Harris}}, \bibinfo {author} {\bibfnamefont {R.}~\bibnamefont {Linford}}, \bibinfo {author} {\bibfnamefont {M.}~\bibnamefont {Mauel}}, \bibinfo {author} {\bibfnamefont {K.}~\bibnamefont {McCarthy}}, \bibinfo {author} {\bibfnamefont {D.}~\bibnamefont {Meade}}, \bibinfo {author} {\bibfnamefont {F.}~\bibnamefont {Najmabadi}},  \emph {et~al.},\ }\bibfield  {title} {\enquote {\bibinfo {title} {Priorities, gaps and opportunities: Towards a long-range strategic plan for magnetic fusion energy},}\ }\href@noop {} {\bibfield  {journal} {\bibinfo  {journal} {US Department of Energy Fusion Energy Sciences Advisory Committee Report No. DOE/SC-0102}\ } (\bibinfo {year} {2007})}\BibitemShut {NoStop}%
\bibitem [{\citenamefont {Andruczyk}\ \emph {et~al.}(2020)\citenamefont {Andruczyk}, \citenamefont {Maingi}, \citenamefont {Kessel}, \citenamefont {Curreli}, \citenamefont {Kolemen}, \citenamefont {Canik}, \citenamefont {Pint}, \citenamefont {Youchison},\ and\ \citenamefont {Smolentsev}}]{andruczyk2020domestic}%
  \BibitemOpen
  \bibfield  {author} {\bibinfo {author} {\bibfnamefont {D.}~\bibnamefont {Andruczyk}}, \bibinfo {author} {\bibfnamefont {R.}~\bibnamefont {Maingi}}, \bibinfo {author} {\bibfnamefont {C.}~\bibnamefont {Kessel}}, \bibinfo {author} {\bibfnamefont {D.}~\bibnamefont {Curreli}}, \bibinfo {author} {\bibfnamefont {E.}~\bibnamefont {Kolemen}}, \bibinfo {author} {\bibfnamefont {J.}~\bibnamefont {Canik}}, \bibinfo {author} {\bibfnamefont {B.}~\bibnamefont {Pint}}, \bibinfo {author} {\bibfnamefont {D.}~\bibnamefont {Youchison}}, \ and\ \bibinfo {author} {\bibfnamefont {S.}~\bibnamefont {Smolentsev}},\ }\bibfield  {title} {\enquote {\bibinfo {title} {A domestic program for liquid metal pfc research in fusion},}\ }\href@noop {} {\bibfield  {journal} {\bibinfo  {journal} {Journal of Fusion Energy}\ }\textbf {\bibinfo {volume} {39}},\ \bibinfo {pages} {441--447} (\bibinfo {year} {2020})}\BibitemShut {NoStop}%
\bibitem [{\citenamefont {Horacek}\ \emph {et~al.}(2021)\citenamefont {Horacek}, \citenamefont {Cecrdle}, \citenamefont {Tskhakaya}, \citenamefont {Dejarnac}, \citenamefont {Schwartz}, \citenamefont {Komm}, \citenamefont {Cavalier}, \citenamefont {Adamek}, \citenamefont {Lukes}, \citenamefont {Veselovsky} \emph {et~al.}}]{horacek2021predictive}%
  \BibitemOpen
  \bibfield  {author} {\bibinfo {author} {\bibfnamefont {J.}~\bibnamefont {Horacek}}, \bibinfo {author} {\bibfnamefont {J.}~\bibnamefont {Cecrdle}}, \bibinfo {author} {\bibfnamefont {D.}~\bibnamefont {Tskhakaya}}, \bibinfo {author} {\bibfnamefont {R.}~\bibnamefont {Dejarnac}}, \bibinfo {author} {\bibfnamefont {J.}~\bibnamefont {Schwartz}}, \bibinfo {author} {\bibfnamefont {M.}~\bibnamefont {Komm}}, \bibinfo {author} {\bibfnamefont {J.}~\bibnamefont {Cavalier}}, \bibinfo {author} {\bibfnamefont {J.}~\bibnamefont {Adamek}}, \bibinfo {author} {\bibfnamefont {S.}~\bibnamefont {Lukes}}, \bibinfo {author} {\bibfnamefont {V.}~\bibnamefont {Veselovsky}},  \emph {et~al.},\ }\bibfield  {title} {\enquote {\bibinfo {title} {Predictive modelling of liquid metal divertor: from compass tokamak towards upgrade},}\ }\href@noop {} {\bibfield  {journal} {\bibinfo  {journal} {Physica Scripta}\ }\textbf {\bibinfo {volume} {96}},\ \bibinfo {pages} {124013} (\bibinfo {year} {2021})}\BibitemShut {NoStop}%
\bibitem [{\citenamefont {Mao}\ \emph {et~al.}(2018)\citenamefont {Mao}, \citenamefont {Fursdon}, \citenamefont {Chang}, \citenamefont {Zhang}, \citenamefont {Liu}, \citenamefont {Ellwood}, \citenamefont {Qian}, \citenamefont {Qin}, \citenamefont {Peng},\ and\ \citenamefont {Barrett}}]{mao2018exploring}%
  \BibitemOpen
  \bibfield  {author} {\bibinfo {author} {\bibfnamefont {X.}~\bibnamefont {Mao}}, \bibinfo {author} {\bibfnamefont {M.}~\bibnamefont {Fursdon}}, \bibinfo {author} {\bibfnamefont {X.}~\bibnamefont {Chang}}, \bibinfo {author} {\bibfnamefont {J.}~\bibnamefont {Zhang}}, \bibinfo {author} {\bibfnamefont {P.}~\bibnamefont {Liu}}, \bibinfo {author} {\bibfnamefont {G.}~\bibnamefont {Ellwood}}, \bibinfo {author} {\bibfnamefont {X.}~\bibnamefont {Qian}}, \bibinfo {author} {\bibfnamefont {S.}~\bibnamefont {Qin}}, \bibinfo {author} {\bibfnamefont {X.}~\bibnamefont {Peng}}, \ and\ \bibinfo {author} {\bibfnamefont {T.}~\bibnamefont {Barrett}},\ }\bibfield  {title} {\enquote {\bibinfo {title} {Exploring the engineering limit of heat flux of a w/rafm divertor target for fusion reactors},}\ }\href@noop {} {\bibfield  {journal} {\bibinfo  {journal} {Nuclear Fusion}\ }\textbf {\bibinfo {volume} {58}},\ \bibinfo {pages} {066014} (\bibinfo {year} {2018})}\BibitemShut {NoStop}%
\end{thebibliography}%

\end{document}